\documentclass[10pt,twocolumn,twoside]{IEEEtran} 

\IEEEoverridecommandlockouts                              

\usepackage{bbold}
\usepackage{upgreek}
\usepackage{color}
\usepackage{graphics} 
\usepackage{needspace}

\input{mysymbol.sty}
\usepackage{theorem}
\usepackage{cite}
\usepackage{amsmath}
\usepackage{amssymb}
\usepackage{mathtools}
\usepackage{multirow}
\usepackage{url}
\usepackage{graphics, subfigure, times, amsfonts}
\usepackage{tikz, epic,eepic}
\usetikzlibrary{shapes,arrows}
\usepackage{pgfplots}
\usepackage{color}
\usepackage{hyperref}
\usepackage{epstopdf}
\usepackage{epsfig, amsbsy}
\usepackage{latexsym}
\usepackage{amscd, verbatim}
\usepackage{multirow}
\usepackage{booktabs}

\usepackage{algorithm}
\usepackage{algorithmic}

 \renewcommand{\baselinestretch}{0.98}\selectfont

\newcommand{\QED}{\hfill\ensuremath{\blacksquare}}

\newtheorem{theorem}{Theorem}

\newtheorem{lemma}{Lemma}
\newtheorem{definition}{Definition}

{\itshape}{\rmfamily}
\newtheorem{assumption}{Assumption}
\newtheorem{remark}{Remark}

\newcommand{\ones}{{\bf{1}}}

\def\forall{\text{for all\ }}

\title{\LARGE Almost Sure Convergence of Networked Policy Gradient over Time-Varying Networks in Markov Potential Games}
\author{Sarper Ayd\i n and Ceyhun Eksin %
\thanks{S. Ayd\i n is with the Department of Industrial and Management Systems Engineering, University of South Florida, Tampa, FL, 33620, USA E-mail: {\tt\small sarperaydin@usf.edu}. C. Eksin is with the Industrial and Systems Engineering Department, Texas A\&M University, College Station, TX 77843. E-mail:{\tt\small eksinc@tamu.edu}%
}}

\begin{document}
\normalsize
\maketitle

%
\begin{abstract}
We propose networked policy gradient play for solving Markov potential games with continuous and/or discrete state-action pairs. 
Agents use parametrized and differentiable policies that depend on the current state and the policy parameters of other agents. \blue{During training, agents update their
policy parameters using stochastic policy gradients while maintaining
estimates of the other agents' parameters through consensus over a
time-varying communication network. We construct an episodic sampling
scheme that yields unbiased policy-gradient estimators. }
\blue{By exploiting the
gradient-alignment property of Markov potential games, we establish almost
sure convergence to a stationary point of the potential value function and an $\mathcal{O}(1/\epsilon^2)$ iteration complexity for reaching an
$\epsilon$-stationary point. Our guarantees accommodate policy score
functions with bounded moments and allow agents to begin with inaccurate and mutually inconsistent estimates of other agents' parameters.} Numerical experiments on a \blue{dynamic multi-agent target assignment problem and a multi-agent newsvendor game} verify the convergence of local beliefs and gradients. \blue{They further show
that networked policies attain higher rewards than communication-free local and independent policies and a centralized-training baseline, while
retaining their performance using time-varying networks with only
$\mathcal{O}(1)$ transmissions per iteration.} 
\end{abstract}

%

%
\section{Introduction}

Markov (stochastic) games describe the rational behavior of selfish decision-makers (agents) with individual rewards, depending on the joint decisions, in dynamic environments \cite{shapley1953stochastic}. Markov games are relevant models of  multi-agent autonomous systems in many settings and their solutions to related canonical problems, e.g., navigation \cite{jia2023rapid}, path planning \cite{muralidharan2019path}, and electricity demand management \cite{narasimha2022multi}. The frequent occurrence of Markov games in such systems prompt the need to design efficient algorithms for solving them. In this paper, we provide a networked multi-agent reinforcement learning (MARL) algorithm for solving Markov potential games.

Markov potential games assume the existence of a potential value function that aligns with the changes in individual value functions which are discounted sum of rewards. If this potential value function is known by an entity that can dictate the players what to do, then the entity can solve the Markov game for a potential maximizing joint policy, treating it as a single-agent MDP, using a standard RL algorithm, e.g., value iteration, policy gradient, and assign individual policies to each agent \cite{littman1994markov,gonzalez2013discrete,zazo2016dynamic,macua2018learning}. In the absence of a centralized entity, if the goal is to attain a maximum of a common objective, e.g., the average reward of the agents, and this objective is commonly known, then the agents can act as a team. In such a setting, the agents can independently collect rewards and update their individual policies to maximize the common objective using a MARL method \cite{zhang2021multi}. 

\blue{A distinct and complementary form of coupling arises when an agent must 
coordinate against the \emph{strategies} the other agents are running, rather 
than react to their realized actions after the fact. The value of conditioning 
on others' strategies is well recognized: an agent that accounts for others' 
strategies coordinates and adapts substantially better than an independent 
learner that treats the other agents as a stationary part of the environment 
\cite{foerster2017learning,he2016opponent}. This is especially pronounced under 
commitment, when agents fix plans in advance and reacting to realized actions is 
too late.  In robotics, mobile robots and autonomous 
vehicles plan their routes and maneuvers against the routing and navigation 
strategies of the rest of the fleet to avoid congestion and deadlock, rather 
than replanning only after collisions materialize 
\cite{mason2023multi}. In supply chains, vendors coordinate replenishment 
timing to meet a shared demand while avoiding simultaneous over-supply, which 
requires anticipating the others' replenishment strategies rather than 
observing their shipments after the fact \cite{liu2025multi}. In each of 
these settings, the information that matters is not the current state or the 
realized actions alone, but the \emph{strategies} the other agents are 
running, which are encoded in their policy parameters.}

In this paper, we introduce a class of networked policy gradient algorithms, in which agents have parameterized policies that not just depend on the common state but also on others' policy parameters. Similar to the single-agent or independent policy gradient \cite{williams1992simple,sutton1999policy,zhang2020global,leonardos2021global}, the policy parameters are updated using local policy gradients which require two distinct sets of information as per the policy gradient theorem: (a) an estimate of the value of the current policy, and (b) score functions of all the local policies (local gradients of the log of the policies). Given the dependence of the local policies on others' policy parameters, agents need to maintain an estimate of all the other agents' policy parameters (even if they are not neighbors) in order to be able to execute their local policies, and compute the local gradients. We design an episodic learning scheme similar to \cite{zhang2020global} that consists of two consecutive episodes with random horizon lengths. In the first episode, agents sample their joint state-action pair to compute the score function, and starting from this sample, they collect discounted rewards through the second episode. After the local gradient updates, agents exchange their local estimates of parameters with their neighbors in a time-varying network, and update their local estimates of other agents' parameters.

We prove the convergence of \textit{networked policy gradient play} to a stationary point of the potential value function of the Markov  game almost surely (Theorem \ref{thm_as_con}) with iteration complexity $\ccalO({1}/{\epsilon^2})$ (Theorem \ref{thm_rate}). These results follow from four intermediate results. First, we show that policy gradients are Lipschitz continuous (Lemma \ref{lemma_lip}) and the episodic sampling provides an unbiased estimation of each agent's policy gradient (Lemma \ref{lem_unb}).  Next, we derive the convergence of local beliefs to the true parameter values in expectation (Lemma \ref{lemma_local_conv}). Lastly, we bound the potential value function changes over consecutive updates (Lemma \ref{lem_iter}). \blue{Numerical experiments on a dynamic multi-agent target assignment problem illustrate networked policies using time-varying networks with only $\mathcal{O}(1)$ transmissions per iteration attain higher rewards in comparison to communication-free local and independent policies, and a centralized-training baseline, namely proximal policy optimization (MAPPO) \cite{yu2022surprising}.}



\vspace{-4pt}
\subsection{Related Literature and Contributions}
\label{sec_lit}
Following the terminology  in static games \cite{shamma2005dynamic}, we use \emph{policy gradient play} to refer to simultaneous policy-gradient
updates in which each agent ascends its own value function.  Distributed
variants using consensus have been studied for monotone games
\cite{tatarenko2020geometric,tan2024timestamp} and min--max problems
\cite{liu2022recursive}.

Independent policy-gradient methods have convergence guarantees for
structured Markov games, including potential and zero-sum games
\cite{zhang2024gradient,leonardos2021global,ding2022independent,
giannou2022convergence,fox2022independent,mao2022improving,
daskalakis2020independent,sun2023provably}.
For finite state and action spaces, these results exploit the game
structure and policy parameterization to relate stationarity to Nash
equilibrium. Direct parameterizations are studied in
\cite{zhang2024gradient,leonardos2021global}, while natural
policy-gradient methods are analyzed in
\cite{fox2022independent,ding2022independent}.
Extensions include linear $Q$-function approximation for infinite
spaces \cite{ding2022independent}, deterministic policies
\cite{giannou2022convergence}, and momentum-based variance reduction
\cite{mao2022improving}.
We instead study policies coupled through agents' parameters over
general, possibly infinite state and action spaces, with convergence
guarantees stated in terms of stationarity.

Other networked MARL methods use consensus to estimate global value functions, $Q$-functions, or rewards \cite{yi2021learning,zhang2018fully,qu2019value}. A separate line combines fictitious play or best-response learning
with reinforcement learning, maintaining beliefs about other agents' policies in finite Markov games
\cite{sayin2022fictitious,baudin2022fictitious}.
Methods for networked MDPs instead exploit local transition structure, where each agent's next state depends on neighboring states and actions, using policy gradients \cite{lin2021multi} or policy iteration \cite{zhang2023global}. Our consensus process tracks other agents' policy parameters for use in coupled policies.

\blue{Our contributions and distinctions from prior work are as follows.

\noindent 1) \textit{Networked Policies and Gradient Estimation:}
We develop policy-gradient play for policies that depend on other agents' parameters, introducing cross-agent score terms absent from independent-policy methods.
We incorporate advantage and temporal-difference estimators and establish unbiasedness (Lemma~\ref{lem_unb}) and Lipschitz continuity (Lemma~\ref{lemma_lip}) under bounded-second-moment assumptions on the score functions. Compared with \cite{zhang2020global,aydin2023networked,aydin2023policyL4DC, aydin2023policy}, we relax almost-sure boundedness and account for baseline cancellation under cross-agent coupling.

\noindent 2) \textit{Consensus under Imperfect Information:}
We establish convergence of agents' beliefs about others' parameters under bounded-moment gradient estimates, admissible polynomially decaying stepsizes, and inaccurate initial beliefs (Lemma~\ref{lemma_local_conv}).
This relaxes the boundedness and initial-agreement assumptions in \cite{eksin2017distributed,arefizadeh2019distributed,
aydin2021decentralized}.

\noindent 3) \textit{Almost-Sure Convergence and Iteration Complexity:} We control the interaction between consensus errors and sampling noise to establish almost-sure convergence of networked policy-gradient play to stationarity of the potential value function (Theorem~\ref{thm_as_con}), strengthening the convergence-in-probability guarantees of
\cite{aydin2023policy,aydin2023networked,aydin2023policyL4DC}.
The method attains $\mathcal{O}(\epsilon^{-2})$ iteration complexity (Theorem~\ref{thm_rate}). For general state and action spaces, stationarity need not imply Nash equilibrium. 
}
\section{Policy Learning in Markov Potential Games}
\label{sec:main}

\subsection{Notation} \label{sec_not}
We denote the Euclidian norm with $||\cdot||$ and the absolute values of scalars and cardinalities of sets with $|\cdot|$.  \blue{All spaces are standard Borel spaces covering countable sets and 
Euclidean spaces.  For a measurable space $(\ccalX,\Sigma_\ccalX)$, $\Delta(\ccalX)$ 
denotes the set of probability measures on it.\footnote{\blue{Unless
otherwise stated, probability measures and stochastic kernels are assumed to be absolutely continuous with
respect to an appropriate $\sigma$-finite reference measure and identify them
with their corresponding densities (e.g., probability mass functions for
counting measure and probability density functions for Lebesgue measure).}}} $\mathbb{1}_{(\cdot)}  \in \{0,1\}$ is an indicator function defined for an event. We define the set of agents as $\ccalN=\{1,\cdots,N\}$, where $|\ccalN|=N$. We use the standard index notation $(i,-i)$ to indicate and differentiate agent $i \in \ccalN$ from the set of the other agents $-i:= \{ j \in \ccalN: j \not = i \}$. Similarly, we identify the ownership of a set $\ccalX$ by using sub-indices such as $\ccalX_i$, $\ccalX_{ij}$ or $\ccalX_{-i}$. We also define the Cartesian product over a set of agents as $\ccalX^N:=\bigtimes_{i \in \ccalN} \ccalX_i$ for a collection of sets $\{\ccalX_i\}_{i \in \ccalN}$. For the notational ease, we denote the sequential sample in the time interval $1 < t< T$ as  $x_{1:T}:=(x_1,\cdots,x_t,\cdots,x_T)$ and denote the estimations of random variables $x$ as $\hat{x}$.

We define the for-loop $\textbf{for} \enspace \tau= t,\cdots,T \enspace \textbf{do}$ as an iterative process of operations over a discrete set of time indices $\{t,\cdots,T\}$ including the last time index $T$. If $t=T$, algorithms only implement the operation once, and if $T<t$, we assume that algorithms skip the for-loop.

 We write $x(i)$ and $X(i,j)$ for the $i^{th}$ entry of a vector $x$ and the $ij$-th entry of a matrix $X$, respectively. We further define the backward product of matrices, defined as follows, for any $T \ge t >0$,
$
    \prod_{\tau=t}^T X_{\tau}:=
 X_T \cdots X_{t+1} X_{t}.
$
\vspace{-4pt}
\subsection{Markov Potential Games} \label{sec_Markov}
We consider a Markov game among a set of finite agents $\ccalN$. Agents take actions from a common action set (without loss of generality) $a_i \in  \ccalA \subseteq \reals^K$ where $K \in \naturals^+$ given a common state $s \in \ccalS$. Note that the set of actions $\ccalA$ and $\ccalS$ do not necessarily have finite cardinality. We denote the joint actions as $a=(a_1,\cdots,a_N) \in \ccalA^N$. \blue{Agents transit between states according to a stochastic kernel $\ccalP$ on 
$\ccalS$ given $\ccalS\times\ccalA^N$, with density $\ccalP^a_{s',s}$, where 
initial states $s_0$ are drawn from $\rho\in\Delta(\ccalS)$.} Agents individually gather rewards $r_i:\ccalS \times \ccalA^N \rightarrow \reals$ at each time $t \in \naturals$, with a discount rate $\gamma \in (0,1)$. We formally state Markov games with the tuple $\Gamma:=(\ccalN,\ccalA^N,\ccalS, \{r_i\}_{i \in \ccalN},\ccalP,\gamma,\rho)$.

\blue{Agents implement policies $\pi_i:\ccalS\to\Delta(\ccalA_i)$, where
$\pi_i(\cdot\mid s)$ is a stochastic kernel on $\ccalA_i$ given $s\in\ccalS$. We use
$\pi_i(a_i\mid s)$ to denote its density whenever convenient.} We define the value function for each agent $i \in \ccalN$,
\begin{equation}\label{eq_val}
V_i^\Pi (s)=\mathbb{E}_\blue{\Pi} \left [ \sum_{t=0}^\infty\gamma^t r_i (s_t,a_t) |s_0=s \right],
\end{equation}
where agents collect rewards with the discount rate $\gamma$ over an infinite horizon as the result of joint policy $\Pi=\bigtimes_{i \in \ccalN} \pi_i$. Note that the expectation operator $\mathbb{E}_\Pi[.]$ is defined over the states and actions distributed with the joint distribution of agents' policies and state-action transitions \footnote{For notational convenience, we will omit the explicit distributional notation under the expectation operator and only include it when necessary.}.  $Q$-function for agent $i$ ($Q_i: \ccalS \times \ccalA^N \rightarrow \reals$) represents the expected sum of discounted rewards given the initial state and action pairs $(s,a) \in \ccalS \times \ccalA^N$,
\begin{equation}\label{eq_Q}
    Q_i^\Pi (s,a)= \mathbb{E}_\blue{\Pi}\left [ \sum_{t=0}^\infty\gamma^t r_i (s_t,a_t) |s_0=s, a_0=a \right].
\end{equation}

Here, we examine the setting where agents use smooth and differentiable 
parameterized policies. \blue{Each agent's 
individual policy $\pi_{i,\theta} (a_i | s):=\pi_i(a_i \mid s; \theta)$ depends on the joint parameter $\theta := (\theta_i, \theta_{-i})$ where the individual policy 
parameters $\theta_i \in \reals^{M_i}$ have dimensions satisfying 
$\sum_{i \in \ccalN} M_i= M$ with $M_i \in \naturals^+$ and $M\in \naturals^+$. 
The joint parameterized policy $\Pi_\theta$ is a product of (conditionally independent) individual policies $\pi_{i,\theta} (a_i| s)$ $\forall{i \in \ccalN}$ given the state and joint policy parameters, i.e.,
\begin{equation}\label{eq_ind_pol}
     \Pi_\theta(a \in \ccalA_q^N|s)=\prod_{i \in \ccalN} \pi_{i,\theta} (a_i \in \ccalA_{i,q}| s)
\end{equation}
%
where $\ccalA_q^N=\bigtimes_{i \in \ccalN} \ccalA_{i,q}$ and $\ccalA_{i,q}$ are measurable subsets over the joint and individual set of actions in order. Note that the policy functions as probability distributions satisfy the standard axioms, i.e., non-negativity, the probabilities of all outcomes sum to 1, and the probability of a union of disjoint outcomes is the sum of their individual probabilities.

The dependence of $i$'s policy $\pi_{i,\theta}$ on the parameters of other players, i.e., $\theta_{-i}$, is the defining feature of the networked policy 
class and encodes the anticipatory coordination, i.e., agent 
$i$'s decision rule reacts to the strategies of the others, represented by 
$\theta_{-i}$, rather than only to the shared state $s$. } It is possible to implement differentiable policies in both finitely or uncountably many actions, (e.g. Gaussian and softmax policies) together with continuous and finite action spaces. We define  differentiable value functions $u_i^{\blue{s}}: \reals^{M} \rightarrow \reals$ given parameterized policies,
\begin{equation}\label{eq_util}
   \hspace{-4pt} u_i^{\blue{s}}(\theta_i,\theta_{-i}):= \mathbb{E}_{\blue{\Pi_\theta}}\Big [ \sum_{t=0}^\infty\gamma^t r_i (s_t,a_t) |s_0=s \Big]. 
\end{equation}

Potential games in static settings assume that there exists a global common objective, i.e., a potential function, that captures utility value changes with respect to unilateral deviations \cite{monderer1996potential}. Markov potential games consider analogous definition for dynamic states assuming this property holds for each state--see, e.g. \cite{macua2018learning}.
We provide an analogous definition of Markov potential games with parametrized policies. 
\begin{definition}[Markov Potential Games] \label{def_mpot}
The game $\Gamma$ is a Markov potential game with individual value functions $u_i$, if there exists a potential value function $\Phi^{\blue{s}}: \reals^M \rightarrow \reals$ \blue{for any fixed initial state $s$} satisfying, 
\begin{equation} \label{eq_mpot}
    \nabla_i u_i^{\blue{s}}(\theta_i, \theta_{-i})=\nabla_i\Phi^{\blue{s}}(\theta) \quad \forall{\theta \in \reals^M}
\end{equation}
where $\nabla_i (.) =\frac{\partial (.)}{ \partial \theta_i}$ is the partial derivative of a given function with respect to the agent $i'$s parameters $\theta_i$.
\end{definition}
\blue{See \cite{arefizadeh2024characterizations,chew2016potential} for results on the existence and construction of potential games, and see \cite{yan2025markovpotentialgameconstruction,liu2025population} for practical applications of Markov potential games. 
Following canonical notation \cite{sutton2018reinforcement,zhang2020global}, we hide the dependence of $u_i$ on state $s$. }
In the rest of the paper, we refer to the parametrized local value functions $u_i$ and $\Phi$ as utility and potential functions respectively. 


\blue{\begin{definition}[Local Nash Equilibrium]\label{def_localne}
A joint parameter profile $\theta^\star=(\theta_1^\star,\dots,\theta_N^\star)$ is a {local Nash equilibrium (NE)} for any fixed initial state 
$s$, if for every agent $i\in\ccalN$, there exists $\epsilon_i>0$ such that
\begin{equation}
u_i(\theta_i^\star,\theta_{-i}^\star)\geq u_i(\theta_i',\theta_{-i}^\star)
\quad\text{for all }\theta_i'\text{ with }\|\theta_i'-\theta_i^\star\| < \epsilon_i,
\end{equation}
that is, no agent can increase its utility by a sufficiently small unilateral deviation.
\end{definition}
 By Definitions~\ref{def_mpot} and \ref{def_localne}, a unilateral deviation improves an agent's utility if and only if it increases the potential. A local NE is thus a coordinate-wise local maximum of the potential, and a Nash equilibrium is the special case in which the neighborhood is the entire parameter space, that is, a coordinate-wise global maximum. Hence every joint local maximum of the potential is a local NE, and every local NE is a stationary point of the potential. Both inclusions are strict when the potential is nonconvex. Local NE requires local maximality along each coordinate block separately, which does not imply local maximality along joint directions. Conversely, a stationary point does not need to be a local NE, since vanishing partial gradients are also consistent with a local minimum or a saddle of some agent's block map. }


\section{Policy Gradient Play with Networked Agents}
\label{sec_learning}
\blue{We state the policy gradient theorem, which expresses the gradient of each 
agent's utility as an expectation involving the value of the current policy 
and the score functions of the policies. }
\vspace{-3pt}
\begin{lemma} \label{lem_pg_def}
The policy gradient of utility function $u_i$ with respect to the parameters $\theta_i$ can be stated as follows,
\begin{align} \label{eq_gradient}
\nabla_i u_i(\theta_i,\theta_{-i}) &= \frac{1}{(1-\gamma)}\mathbb{E}\bigg[ (Q_i^{\Pi_{\theta}} (s,a)-b_i(s)) \nonumber \\
& \quad \times \sum_{n \in \mathcal{N}} \nabla_i \log \pi_{n,\theta} (a_n |s)\big| \theta \bigg]
\end{align}
where $b_i: \ccalS \rightarrow \reals$ is a baseline function for each agent $i \in \ccalN$, independently defined from the joint actions of agents $a \in \ccalA^N$.
\end{lemma}
\vspace{-3pt}


\blue{See Appendix for the proof. As per 
\eqref{eq_gradient}, the local policy gradient of agent $i$ has two components, 
both depending on the joint parameters $\theta=(\theta_i,\theta_{-i})$. The 
\emph{reward term} $Q_i^{\Pi_{\theta}}(s,a)-b_i(s)$ depends on $\theta$ through 
the joint policy governing the collected rewards, while the \emph{joint score} 
$\sum_{n\in\ccalN}\nabla_i\log\pi_{n,\theta}(a_n|s)$ sums the score functions of 
all agents' policies, generalizing the single-agent score of 
\cite{zhang2020global} through the coupled parameterization. 

We can express the reward term using the $Q$-function or the advantage function 
$A_i^{\Pi_{\theta}}(s,a):= Q_i^{\Pi_{\theta}}(s,a)-V_i^{\Pi_{\theta}}(s)$ by 
setting $b_i(s)=0$ or $b_i(s)=V^{\Pi_\theta}_i(s)$, respectively, and 
equivalently through the temporal-difference error 
$TD_i(s,a):=r_{i}(s,a)+\gamma V^{\Pi_{\theta}}_i(s')-V_i^{\Pi_{\theta}}(s)$, 
where $s'$ follows the transition kernel.}

\vspace{-5pt}
\subsection{{Coupled Stochastic Policy Gradient Updates}}
The local parameter updates follow stochastic gradient updates
\vspace{-2pt}
 \begin{equation} \label{eq_pg}
    \theta_{i,t}=\theta_{i,t-1}+\alpha_{t} \hat{\nabla}_i u_i(\theta_{i,t-1},\theta_{-i,t-1}),
\end{equation}
where $\alpha_t \in \reals^+$ is a common decaying step size and $\hat{\nabla}_i u_i(\theta_{i,t-1},\theta_{-i,t-1})$ is the stochastic gradient. Using \eqref{eq_gradient}, we can express the stochastic gradient as follows
\begin{align} 
    \hat{\nabla}_i u_i(\theta_{i},\theta_{-i}) := \frac{1}{(1-\gamma)}\hat{R}_{i} \sum_{n \in \ccalN} \nabla_i \log  \pi_{n,\theta} (a_{n,\ccalT_{1}} |s_{\ccalT_{1}}) \label{eq_def_sg}
\end{align}
where $\hat{R}_i:= \hat{Q}_i-\hat{b}_i$ is the estimator for the reward term in \eqref{eq_gradient} and $(s_{\ccalT_{1}}, a_{n,\ccalT_{1}})$ is a sampled state-action pair over an episode with length $\ccalT_1$. We consider $\hat{Q}^{\Pi_{\theta}}_{i}$, advantage function $\hat{A}^{\Pi_{\theta}}_{i}$ or the temporal difference $\widehat{TD}^{\Pi_{\theta}}_{i}$ for $\hat{R}_i$.

Next, we describe an estimation scheme for these two components. Following the episodic approach of
\cite{zhang2020global}, before each gradient update the agents sample a
horizon $\ccalT_1\sim\operatorname{Geom}(1-\gamma)$ and generate a trajectory to obtain the state--action pair $(s_{\ccalT_1},a_{\ccalT_1})$.\footnote{For $\widetilde{\gamma}\in(0,1]$, we define $\operatorname{Geom}(\widetilde{\gamma})$ on $\{0,1,\ldots\}$ by $\mathbb{P}(\ccalT=x)=(1-\widetilde{\gamma})^x \widetilde{\gamma}$.}
Starting from this pair, they collect rewards over a second random horizon with length $\ccalT_2\sim\operatorname{Geom}(1-\sqrt{\gamma})$ and use their discounted sum to construct $\hat Q_i(s_{\ccalT_1},a_{\ccalT_1})$.

The advantage and temporal-difference estimators additionally use value estimates obtained from independent rollouts of the same random-horizon form, with the initial action sampled from the policy. \blue{Specifically, the advantage estimator combines $\hat Q_i(s_{\ccalT_1},a_{\ccalT_1})$ with $\hat V_i(s_{\ccalT_1})$. The temporal-difference estimator instead combines the immediate reward $r_i(s_{\ccalT_1},a_{\ccalT_1})$ with $\hat V_i(s_{\ccalT_1})$ and $\hat V_i(s_{\ccalT_1+1})$, where $s_{\ccalT_1+1}$ is the next state reached under $a_{\ccalT_1}$; it therefore does not require a separate
$Q$-estimate.} All gradient estimates depend only on observed rewards and
actions and the policy parameters, without requiring knowledge of the
transition kernel $\ccalP^{a}_{s',s}$. Algorithm~\ref{alg_est_inner}
summarizes the episodic gradient-estimation procedure.

\begin{algorithm}[H] 
   \caption{ Gradient Estimation for Agent $i$}
\label{alg_est_inner}
\begin{algorithmic}[1]\label{alg_est_PG}
 \STATE {\bfseries Input:} The parameters 
 $\theta$, initial state $s_0$,  and discount factor $\gamma$.
    \STATE Draw $\ccalT_1 \sim Geom(1-\gamma)$ and reset $s_0$. 
    \STATE Sample action
    $a_{i,0} \sim \pi_{i,\theta}(.|s_{0})$ \forall $i\in \ccalN$
    \FOR{$\tau=0,1,\cdots, \ccalT_1-1$} 
    \STATE Reach state $s_{\tau+1} \sim \ccalP^{a_{\tau}}_{s_{\tau+1},s_{\tau}}$
    \STATE Sample and take action, 
    $a_{i,\tau+1} \sim \pi_{i,\theta}(.|s_{\tau+1})$  
    \ENDFOR
    \STATE Compute $\nabla_i \log \pi_\theta(a_{\ccalT_1}|s_{\ccalT_1} )$ \forall $i\in\ccalN$
\STATE \blue{Draw $\ccalT_2 \sim Geom(1-\gamma^{0.5})$ and set 
       $\hat Q_i = \sum_{t=\ccalT_1}^{\ccalT_1+\ccalT_2}\gamma^{(t-\ccalT_1)/2} r_i(s_t,a_t)$}
\STATE \blue{Form $\hat V_i(s)$ by an independent rollout as in line~9 from $s$, 
       with the initial action drawn from $\pi_{i,\theta}(\cdot|s)$}
\STATE \blue{Set $\hat R_i = \hat Q_i$, \; $\hat Q_i - \hat V_i(s_{\ccalT_1})$, \; or \; 
       $r_i(s_{\ccalT_1},a_{\ccalT_1}) + \gamma\hat V_i(s_{\ccalT_1+1}) - \hat V_i(s_{\ccalT_1})$}
    \STATE Return $\hat{\nabla}_i u_i(\theta_i,\theta_{-i})$ using \eqref{eq_def_sg}

\end{algorithmic}
\end{algorithm}
\blue{\begin{remark}
Executing Algorithm~\ref{alg_est_PG} requires two types of coordination. 
First, agents must act upon the common random horizons $\ccalT_1$ and 
$\ccalT_2$, which is resolved by fixing a common random seed before training. 
Second, the score functions of joint policies take the joint actions as an 
input, and we assume the realized action profile is observed at $\ccalT_1$. 
This is a single observation of $a_{\ccalT_1}$ per episode rather than continuous tracking of others' actions. This joint action profile can be available in 
scenarios such as drone navigation~\cite{hodge2021deep}, collision 
avoidance~\cite{han2022deep}, and vehicle tracking~\cite{li2020vehicle}. 
Otherwise the action must be relayed across the network before the gradient is 
formed, or replaced by a sample from agent $i$'s belief 
$\pi_{n,\hat\theta^i_n}(\cdot|s_{\ccalT_1})$. We leave these partially observed regimes 
as open directions.
\end{remark}}
\subsection{Beliefs, Information Exchange and Consensus}
Agents need to know the values of other agents' parameters $\theta_{-i,t}$ in order to compute policies \eqref{eq_ind_pol} and stochastic gradients \eqref{eq_def_sg}. This information may not be available during training. Instead, we assume agents keep local beliefs $\hat{\theta}^i_{-i,t}$ on $\theta_{-i,t}$. Agents form these beliefs by exchanging their beliefs with their neighbors $\ccalN_{i,t}:= \{j: (i,j)\in {\ccalE_t}\}$ over time-varying communication networks $\ccalG_t=(\ccalN,{\ccalE_t})$. 
The update on agent $i$'s belief on agent $j$'s $m$th parameter $\hat{\theta}_{j,t}^i(m) \in \reals$ follows a consensus update 
\begin{equation}\label{eq_local_update}
    \hat{\theta}_{j,{t+1}}^i(m)= \sum_{l \in \ccalN_{i,t} \bigcup \{i\} }w_{jl,t}^i \hat{\theta}_{j,t}^{l}(m),
\end{equation} 
where $w^i_{jl,t} \ge 0$ is the weight of agent $i$ on agent $l$'s estimate of agent $j$'s $m^{th}$ parameter.
 We make the following assumptions on the communication network and the weights.
\begin{assumption}\label{as_connect}
The network $\ccalG=(\ccalN,\ccalE_{\infty})$ is connected, where $\ccalE_{\infty}=\{(i,j)| (i,j) \in \ccalE_t, \, \text{for infinitely many t} \in \naturals \}$. There exists a time step $T_{\ccalG}>0$, such that for any edge $(i,j) \in \ccalE_{\infty}$ and $t \ge 1$, it holds  $(i,j) \in \bigcup_{\tau=0}^{T_{\ccalG}-1}\ccalE_{t+\tau}$.
\end{assumption}
This assumption, first introduced in \cite{NedicOzdaglar}, assures that any pair of agents are connected after some finite time, and any edge $(i,j)\in \ccalE_{\infty}$ appears in a bounded time interval {$T_\ccalG$}. 
%
\begin{assumption}\label{as_com_weights}
There exists a scalar $h \in (0,1)$, such that the following statements hold for all $i\in \ccalN$, $j\in \ccalN$ and $t \in \naturals^+$,
\begin{itemize}
    \item[\textit{(i)}] If $l \in \ccalN_{i,t} \cup  \{i\}$, then $w_{jl,t}^i \ge h$, else  $w_{jl,t}^i=0$,
    \item[\textit{(ii)}] $w_{jj,t}^j =1$,  
    \item [\textit{(iii)}] $\sum_{l \in \ccalN_{i,t} \cup \{i\} } w_{jl,t}^i=1$.
\end{itemize}
\end{assumption}
Assumption \ref{as_com_weights}{\it (i)} means that agents only assign positive weights to their current neighbors and themselves based on the edges $\ccalE_t$. Assumption \ref{as_com_weights}{\it (ii)} means that agents do not utilize any information from other agents to form a belief on their own parameters such that it holds $\hat{\theta}^j_{j,t}=\theta_{j,t}$ for all $j \in \ccalN$ and $t\geq 0$. 
Assumption \ref{as_com_weights}{\it(iii)} implies that the weight matrices are row stochastic.
\blue{
\begin{remark}
    The consensus update in Eq.~ \eqref{eq_local_update} requires each agent to store an additional belief vector 
$\hat\theta^i_{-i,t}\in\reals^{M-M_i}$ on all other agents' parameters, a 
storage cost of order $\ccalO(M)$ per agent. Communication, however, is 
neighbor-to-neighbor: the consensus update  exchanges 
beliefs only within $\ccalN_{i,t}\cup\{i\}$, so the load per iteration is 
$\ccalO(|\ccalE_t|\cdot M)$. Assumption~\ref{as_connect} allows $|\ccalE_t|$ to be as small as 
a single edge per iteration.
\end{remark}}
The following assumption allows for bounded deviations in the initial belief error in expectation, rather than requiring that all initial beliefs are correct as in prior studies \cite{eksin2017distributed,arefizadeh2019distributed,aydin2021decentralized,aydin2023networked,aydin2023policy,aydin2023policyL4DC}. 

\begin{assumption}[Bounded Deviation of Initial Belief Errors]  \label{as_belief_error} 
For any pair of agents $(i,j) \in \mathcal{N} \times \big(\mathcal{N} \setminus \{i\}\big)$ and any parameter $\theta_{j,0}(m) \in \mathbb{R}$ of agent $j$, the first moment of the initial belief error is bounded, i.e.,  
$\mathbb{E}[|\hat{\theta}^i_{j,0}(m)-\theta_{j,0}(m)|] \le \kappa$ where $\kappa\ge0$.
\end{assumption}

\subsection{Networked policy gradient play}
Networked policy gradient play (Algorithm \ref{alg_DFP}) has three main steps. Agents estimate the gradient direction by playing in two random episodes with lengths $\ccalT_1$ and $\ccalT_2$. Then, they update their parameters using stochastic gradient update. In the last step, agents exchange their parameters with their current neighbors and update their beliefs.

\label{sec:alg}
\begin{algorithm}[H] 
   \caption{ Networked Policy Gradient Play}
\label{suboptimal_alg_inner}
\begin{algorithmic}[1]\label{alg_DFP}

   \STATE {\bfseries Input:} Local estimates $\hat{\theta}_{-i,0}^i$ and $\ccalG_t=(\ccalN,\ccalE_t)$, initial state $s_0$ and initial policy $\Pi_{\theta,0}$, and discount factor $\gamma$.
\FOR{$t=1,2\cdots,$} 
    \STATE Run Algorithm \ref{alg_est_PG} with local beliefs $\hat{\theta}^i_{-i,t}$ for all $i\in \ccalN$
    \STATE Update parameters $\theta_i$  with stochastic gradients $\hat{\nabla}_i u_i(\theta_i,\hat{\theta}_{-i})$ for $i\in \ccalN$ \eqref{eq_pg}.
    \STATE Update local copies $\hat{\theta}_{j,t}^i$ for $j\in -i$ and $i\in \ccalN$ \eqref{eq_local_update}.
  \ENDFOR 
   \end{algorithmic}
\end{algorithm}

\section{Convergence of Networked Policy Gradient Play in Markov Potential Games} \label{sec::conv}
We assume the gradient stepsizes are chosen such that their sum is divergent while the sum of squared stepsizes is convergent, typical for stochastic first-order methods \cite{bottou2018optimization}.
\begin{assumption}\label{as_step}
\blue{The step size $\alpha_t$ has the following decaying form $\alpha_t=\Theta(1/t^\beta)$, where $\beta \in (1/2,1)$}. 
\end{assumption}
%
 The following assumptions are set to satisfy regularity conditions on rewards and policy functions.
\begin{assumption}\label{as_bound_rew}
The reward for any agent $i$ at any state and joint action profile $(s,a) \in \ccalS \times \ccalA^N$ is bounded, $|r_{i}(s,a)| \le R$ where $R>0$. 
\end{assumption}
The bounded reward assumption is commonly used in infinite horizon problems with a discount rate. This ensures that individual utilities and the potential function are also bounded. 

Next, we formally define a probability space $(\Omega, \mathcal{F}, P)$
where the total sample space $\Omega$ is the Cartesian product of individual sample spaces of the parameters $\theta $, beliefs $\hat{\theta}$, episode lengths ($\ccalT_1$ and $\ccalT_2$), actions $a$ and states $s$, i.e,
$ \Omega = \reals^M \times \reals^{MN} \times \naturals \times \naturals \times \ccalS \times \ccalA^N$. \blue{We define the filtration $\{\mathcal{F}_t\}_{t\ge 0}$ as the natural sequence of increasing $\sigma$-algebras 
 \begin{align} \label{eq_filtration}
     \ccalF_t=\sigma(&\{\theta_\tau\}_{\tau \in \{0,\cdots, t\} }, \{\hat{\theta}_{-i,\tau}\}_{(i,\tau) \in \ccalN \times \{0,\cdots, t\}}, \nonumber \\&\{\ccalT_{1,\tau}\}_{\tau \in \{1,\cdots, t-1\}}, \{\ccalT_{2,\tau}\}_{\tau \in \{1,\cdots, t-1\}}, \nonumber 
     \nonumber \\
     &\{(s_{0,\tau},\cdots, s_{\ccalT_1+\ccalT_2+1},\tau)\}_{\tau \in \{1,\cdots, t-1\}}, \nonumber \\
    &\{(a_{0,\tau},\cdots, a_{\ccalT_1+\ccalT_2+1,\tau})\}_{\tau \in \{1,\cdots, t-1\}}).
 \end{align}} 
\begin{assumption}\label{as_bound_logp}
The first and second moments of policy score function of each agent $n\in\ccalN$ with respect to agent $i \in \ccalN$'s parameters exists and is bounded, $\mathbb{E} [||\nabla_i \log \pi_{n,\theta} (a_{n}| s)|| | \theta] \le B$ and, $\mathbb{E} [||\nabla_i \log \pi_{n,\theta} (a_{n}| s)||^2 |\theta] \le V^2$ for any $\theta \in \reals^M$, state $s \in \ccalS$ and action $a_i \in \ccalA$, where $B \ge 0$ and $V \ge 0$. Furthermore, it is Lipschitz continuous, i.e.,  $||\nabla_i \log \pi_{n,\theta^1} (a_{n}| s) -\nabla_i \log \pi_{n,\theta^2}(a_{n}| s)|| \le \ccalL||\theta^1- \theta^2||$ for any $n \in \ccalN$ and $\theta^1, \theta^2 \in \reals^M$, where $\ccalL\ge0$.
\end{assumption}

\begin{remark} \label{rem_gaussian}
    We only require boundedness of the first and second moments of policy score functions. This is a relaxation of the bounded score functions assumption used in former studies on single-agent and networked policy gradient algorithms \cite{zhang2020global,aydin2023policy}.
   While softmax policies have bounded score functions almost everywhere, Gaussian policies do not in general. For example,  suppose that agents implement linearly parametrized mean $\mu_i:=\theta^T\Psi(s)$ of Gaussian policies $\pi_{i,\theta} (a_i|s):= N(\mu_i,\sigma^2)$, where $\Psi(s): \ccalS \rightarrow \reals^M$ are features derived from states of the game. Given the score function $\mathbb{E}[(a_i-\theta^T\Psi(s))\Psi(s)/ \sigma^2]$, three conditions have to be satisfied to have bounded score functions almost surely: \textit{i)} the action set is bounded, \textit{ii)} the norm of features derived from state space is bounded, and \textit{iii)} the iterates of parameters stay in a bounded set. However, stochastic gradient updates with non-summable stepsizes cannot ensure that the gradients stay in bounded regions. For the first bounded moment assumption, it is enough to have the variance of the policy $\sigma^2$ set to a non-zero value if it is a constant value and bounded state features. The reason is that the mean absolute deviation of the sampled action is always less than the standard variation via the Jensen's inequality, i.e. $\mathbb{E}[|a_i-\theta^T\Psi(s)|]/\sigma <1$. For the second moment, bounded features on the state are sufficient along the same lines. 
\end{remark}

%

We state our main convergence results in the following.
\begin{theorem} \label{thm_as_con}
\blue{Suppose Assumptions~\ref{as_connect}--\ref{as_bound_logp} hold. Then, for any
fixed initial state $s_0$, the sequence of joint parameters
$\{\theta_t\}_{t\geq 0}$ generated by Algorithm~\ref{alg_DFP},  
converges almost surely to the set of stationary points of the potential
function $\Phi(\theta)$, that is,
$
\lim_{t\rightarrow\infty}
 \nabla\Phi(\theta_t)=0
$ almost surely.
}
\end{theorem}

The proof is available in Appendix~\ref{ap_proof_as}. See the discussion following Definition \ref{def_localne} for the relation between stationary points and local NE.
Next, we establish the iteration complexity.
\begin{theorem} \label{thm_rate}
Suppose Assumptions~\ref{as_connect}-\ref{as_bound_logp} hold.  
Let $\{\theta_t\}_{t \ge 0}$ be the sequence of parameters updated by Algorithm~\ref{alg_DFP}.  
Define the number of iterations required for $\{\theta_t\}_{t \ge 0}$ to reach an $\epsilon$-stationary point of the potential in expectation \blue{for any fixed initial state $s_0$} as,
\begin{equation}
        T_{\epsilon} := \min \left\{ t \in \mathbb{N} : \inf_{0 \le \tau \le t} \, \mathbb{E}\big[\|\nabla\Phi(\theta_\tau)\|^2\big] \le \epsilon \right\}.
\end{equation}
Then it holds that $T_{\epsilon} = \mathcal{O}(\epsilon^{-1/\tilde{\beta}})$, where $\tilde{\beta} := \min\{1-\beta, \beta\}$.  
In particular, as \blue{$\beta \to 1/2^+$}, the optimal rate is achieved, i.e., $T_{\epsilon} = \mathcal{O}(\epsilon^{-2})$.
\end{theorem}
The proof is provided in Appendix~\ref{app_proof_rate}. Theorem \ref{thm_rate} establishes the $\ccalO(1/\sqrt{t})$ convergence rate for decreasing stepsizes with $\beta \to 1/2^+$, matching the rate of standard stochastic gradient descent in nonconvex optimization \cite{ghadimi2013stochastic,shapiro2021lectures} and in single-agent policy gradient \cite{zhang2020global}.

\subsection{Key Intermediate Results}
Theorems \ref{thm_as_con} and \ref{thm_rate} rely on intermediate results establishing continuity, boundedness and unbiasedness of the gradients (Lemma \ref{lemma_lip}-\ref{lem_unb}), and the consensus rate of the beliefs (Lemma \ref{lemma_local_conv}). We highlight and discuss the novel elements in proof techniques whenever they become relevant.  

\begin{lemma}[Lipschitz and Bounded Policy Gradients] \label{lemma_lip}
Suppose Assumptions \ref{as_bound_rew}-\ref{as_bound_logp} hold. The policy gradient of any agent $i\in \ccalN$,  $\nabla_{i} u_{i} (\theta_{i}, \theta_{-i})$ is Lipschitz continuous with some constant $L>0,$ i.e., for any $\theta^{1}, \theta^{2} \in \mathbb{R}^{M}$ 
\begin{equation}
 ||\nabla_{i} u_{i} (\theta_{i}^1, \theta_{-i}^1) -\nabla_{i} u_{i} (\theta_{i}^{2}, \theta_{-i}^{2})|| \le L ||\theta^{1} -\theta^{2}||,
\end{equation}where the value of the Lipschitz constant $L$ is defined as,
\begin{equation}\label{eq_lip_const}
L:= N R \Bigg (\frac{1}{(1-\gamma)^2} \ccalL +\frac{(1+\gamma)V^2}{(1-\gamma)^3}\Bigg).
\end{equation}
In addition, the policy gradient of any agent $i\in \ccalN$, is bounded, i.e, $ ||\nabla_i u_i(\theta_i,\theta_{-i})|| \le l$, where
$
l: = \frac{2NRB}{(1-\gamma)^2}.$ 

\end{lemma}
See Appendix for the proof. We note that this proof departs from the standard arguments used to establish continuity of the gradient, as it relies on weaker assumptions. In particular, instead of assuming boundedness of the gradient, we only require its moments to be bounded (Assumption \ref{as_bound_logp}).
\begin{lemma} \label{lem_unb}
Suppose Assumptions \ref{as_bound_rew}-\ref{as_bound_logp} hold. Then the stochastic policy gradient $\hat{\nabla}_i u_i(\theta_i,\theta_{-i})$ is unbiased, i.e., $\mathbb{E}[\hat{\nabla}_i u_i(\theta_i,\theta_{-i})|\theta]=\nabla_i u_i(\theta_i,\theta_{-i})$ for $i \in \ccalN$.
\end{lemma}
Proof follows standard arguments in \cite{zhang2020global} or see Appendix D. 
Together, these results ensure that the policy gradients get closer to the true ascent direction as local beliefs converge to true parameter values. The results also guarantee that the changes in parameters stay bounded thanks to bounded policy gradients under the expectation operator.

\begin{lemma} [Consensus of Beliefs]\label{lemma_local_conv}
Suppose Assumptions~\ref{as_connect}-\ref{as_bound_logp} hold. Then, local copies  $\hat{\theta}^i_{j,t}$  converge to $\theta_{j,t}$ with the rate $\ccalO(t^{-\beta})$ in expectation, i.e. $\mathbb{E}[|| \hat{\theta}^i_{j,t}- \theta_{j,t}||]= \ccalO(t^{-\beta})$ for $i\in\ccalN$ and $j\in \ccalN$.  
 \end{lemma}
\begin{IEEEproof}
Let $W_{j,t} \in \mathbb{R}^{N \times N}$ be the weight matrix formed by the scalar entries $ W_{j,t}(i,l)=w^i_{jl,t}$ at time $t$. Using $W_{j,t}$ we can express \eqref{eq_local_update} the updates for $\hat{\theta}_{j,t}(m) \in \mathbb{R}^N$ as
 \begin{equation}\label{eq_rec_main}
     \hat{\theta}_{j,t}(m)=W_{j,t}(\hat{\theta}_{j,t-1}(m)+ (\theta_{j,t}(m)-\theta_{j,t-1}(m))e_j),
 \end{equation}
 where $e_j$ is the canonical vector of $j^{th}$ base in $\mathbb{R}^N$. The second term in \eqref{eq_rec_main} makes it explicit that agent $j$ is a source node in the communication network feeding their updated parameters $\theta_{j,t}$ to the other agents. 

We subtract the vector $\theta_{j,t}(m)\ones$, where $\ones \in \mathbb{R}^N$ is the vector of ones in $\reals^N$, from both sides of \eqref{eq_rec_main}, to obtain,
\begin{align}
\label{eq_rec_2}
&\hat{\theta}_{j,t}(m)-\theta_{j,t}(m)\mathbf{1} \\
\nonumber=
& W_{j,t} (\hat{\theta}_{j,t-1}(m)+(\theta_{j,t}(m)-\theta_{j,t-1}(m))e_j) -\theta_{j,t}(m)\mathbf{1}.
\end{align}
 The term $\theta_{j,t}(m)\ones$ can go inside the matrix multiplication since $W_{j,t}$ is a row-stochastic matrix by Assumption \ref{as_com_weights}. By letting $y_{t}= \hat{\theta}_{j,t}(m)-\theta_{j,t} (m) \mathbf{1}$, we rearrange \eqref{eq_rec_2} as follows, 
   \begin{equation}\label{eq_rec_3}
     y_{t}=W_{j,t}(y_{t-1}+ (\theta_{j,t}(m)-\theta_{j,t-1}(m))e_j-(\theta_{j,t}(m)-\theta_{j,t-1}(m))\ones)\nonumber.
 \end{equation}
 Next, we define and derive an upper bound for the term $\delta_{t}= (\theta_{j,t}(m)-\theta_{j,t-1}(m))(e_j -\ones)$ for $t\ge1$, 
 \blue{ \begin{align}
    ||\delta_t||
    &\le ||(\theta_{j,t}(m)-\theta_{j,t-1}(m))|| (||e_j||+||\ones||) \nonumber \\
    \le& \alpha_t \big|\big|\big[\hat{\nabla}_j u_j(\theta_{j,t-1},\hat{\theta}_{-j,t-1})\big]_m\big|\big|(N+1),
 \end{align}
 where $\big[\hat{\nabla}_j u_j(\theta_{j,t-1},\hat{\theta}_{-j,t-1})\big]_m$ is the $m$-th component of the stochastic policy gradient \eqref{eq_def_sg} of agent $j$'s value function with respect to its parameter at index $m$.}
Since the expected score functions given parameters and rewards of any agent are bounded by Assumptions \ref{as_bound_rew}-\ref{as_bound_logp}, the unconditional expectation of partial gradients is also bounded. Then, we bound the difference $\delta_t$ using Assumption \ref{as_step},
 \blue{ \begin{align}
    \mathbb{E} [ || \delta_t || ] \le&   \alpha_t (N+1)  \mathbb{E} \Big[\big|\big|\big[\hat{\nabla}_j u_j(\theta_{j,t-1},\hat{\theta}_{-j,t-1})\big]_m\big|\big| \Big] \nonumber \\
    &\le \alpha_t (N+1) l =\ccalO(1/t^\beta).\nonumber
 \end{align}}
 Now we rewrite and expand Eq. \eqref{eq_rec_3} as follows,
 \begin{align}
     y_{t}&= W_{j,t}(y_{t-1}+\delta_t)\nonumber\\
     &= \left(\prod_{\uptau=1}^{t} W_{j,\uptau} \right ) y_0 +\sum_{\zeta=1}^{t} \left(\prod_{\uptau=1}^{\zeta} W_{j,t-\uptau+1} \right) \delta_{t-\zeta+1}. \nonumber
 \end{align}
We bound the expected norm of the difference $y_t$ by using the facts that $e_j^T \delta_t = 0$ and $e_j^T y_t = 0$ for $t \in \mathbb{N}^+$. These follow from the fact that $\delta_t(j) = 0$ by definition and $y_t(j) = \hat{\theta}^j_{j,t}(m) - \theta_{j,t}(m) = 0$ by construction. Hence, we have
  \begin{align} \label{eq_norm_y}
      \mathbb{E} [ || y_t || ]
      &\le \left\|\prod_{\uptau=1}^{t} W_{j,\uptau}-\ones e_j^T \right \| \cdot  \mathbb{E} [ || y_0 || ]\nonumber \\
      &+\sum_{\zeta=1}^{t} \left \|\prod_{\uptau=1}^{\zeta} W_{j,t-\uptau+1} -\ones e_j^T \right \| \cdot\mathbb{E} \big[ \big|\big| \delta_{t-\zeta+1}\big|\big|\big]. \nonumber
\end{align}
Using Lemma 1 from \cite{arefizadeh2019distributed},  the backward product of the weight matrices converges to the matrix $\ones e_j^T$, i.e., $\lim_{t \rightarrow \infty} \prod_{\uptau=1}^{t} W_{j,\uptau}=\ones e_j^T$ by Assumptions \ref{as_connect}-\ref{as_com_weights}. Moreover, the convergence rate is geometric, i.e., $||(\prod_{\uptau=1}^{\zeta} W_{j,t-\uptau+1})-\ones e_j^T|| \le \ccalO(\lambda^\zeta)$ with some $\lambda <1 $ for any $\zeta \in \{1,\cdots,t\}$. Therefore,  the norm of each term $||((\prod_{\uptau=1}^{\zeta} W_{j,t-\uptau+1})-\ones e_j^T) || \cdot  \mathbb{E} [ ||\delta_{t-\zeta+1} || ]$ is bounded. For the same reason, and by Assumption \ref{as_belief_error} we also obtain a geometrically decaying effect of the initial belief error, $\left\|\prod_{\uptau=1}^{t} W_{j,\uptau}-\ones e_j^T \right \| \cdot  \mathbb{E} [ || y_0 || ] =\ccalO (\lambda^t) N \kappa= \ccalO (\lambda^t)$, in expectation,
\begin{equation}
 \mathbb{E} [ ||y_{t}|| ] \le \ccalO (\lambda^t)+\sum_{\zeta=1}^{t} \ccalO(\lambda^\zeta) \mathbb{E} [  ||\delta_{t-\zeta+1}||]. \nonumber
\end{equation}
By Chebychev's sum inequality, it holds,
\begin{equation}
 \mathbb{E} [ ||y_{t}|| ] \le  \ccalO (\lambda^t)+ \mathbb{E} [ \delta^{avg}_{t} ] \sum_{\zeta=1}^{t} \ccalO(\lambda^\zeta), \nonumber
\end{equation}
where $\delta^{avg}_{t}=t^{-1}\sum_{\zeta=1}^{t}  \mathbb{E} [ ||\delta_{t-\zeta+1}|| ] $. Then, we have
\begin{equation}
\delta^{avg}_{t}=t^{-1}\sum_{\zeta=1}^{t}||\delta_{t-\zeta+1}||= \frac{\alpha_0 l}{t} \sum_{\zeta=1}^{t}\frac{N+1}{(t-\zeta+1)^{\beta}}=\ccalO(t^{-\beta})  
\end{equation}
Given $\lambda <1$, it yields $\mathbb{E}[||y_{t+1}||]=\ccalO(t^{-\beta})$ as the geometric sum is convergent, and $\sum_{\zeta=1}^{t} \ccalO(\lambda^\zeta) = \ccalO(1/1-\lambda)$.
\end{IEEEproof}
This lemma shows that local beliefs converge to the true values of agents' parameters. This result generalizes the previous result stated in \cite{aydin2023policy} for any stepsize with the parameter $\beta \in (0.5, 1)$ and the case without almost surely bounded policy score functions. Note that we recover the $\ccalO(\log t/t)$ rate established in \cite{aydin2023policy} when $\beta=1$. 

Next, we analyze the expected potential change at each iteration conditioned on the current information available.
 \begin{lemma}\label{lem_iter}
Suppose Assumptions \ref{as_connect}-\ref{as_bound_logp} hold. Let $\Phi_{\max} := \sup_{\theta \in \reals^M}\Phi(\theta)$, and $Z_t=\Phi_{\max} -\Phi(\theta_t)$ be the optimality gap with respect to the parameter values $\theta_t$ at time $t$. The following relation holds between any consecutive time steps $t$ and $t+1$ 
\begin{align} \label{eq_pot_change}
    &\mathbb{E}[Z_{t+1}|\ccalF_{t}]-Z_t \le - \alpha_t ||\nabla\Phi(\theta_t)||^2  \\
    &+\alpha_t \sqrt{N} l L \sum_{i \in \ccalN} \sum_{j \in \ccalN \setminus \{i\}} ||\theta_{j,t}- \hat{\theta}^i_{j,t}||+\frac{2 L N^{3.5} R^2 V^2 }{(1-\gamma)^2(1-\sqrt{\gamma})^2}\alpha^2_t.\nonumber
\end{align}
where $\mathbb{E}[.|\ccalF_{t}]$ is the expectation conditioned on the filtration $\ccalF_{t}$. 
Then, the infinite sum of the product of stepsizes and square norm of gradients is convergent, i.e.,
$    \sum_{t=0}^\infty \alpha_t ||\nabla\Phi(\theta_t)||^2 < \infty.
$%

\end{lemma}

\begin{IEEEproof}
By Taylor's Expansion and Lemmas \ref{lemma_lip}-\ref{lem_unb}, and noting that $\nabla\Phi$ is $\sqrt{N}L$-Lipschitz since it stacks $N$ blocks each $L$-Lipschitz by Lemma \ref{lemma_lip}, we have the relation,
\begin{align}
   \Phi(\theta_{t+1})-\Phi(\theta_{t}) &\ge (\theta_{t+1}-\theta_{t})^T \nabla\Phi(\theta_t)-\frac{1}{2}\sqrt{N} L||\theta_{t+1}-\theta_{t}||^2,\nonumber \\
    & \ge \alpha_t g(\hat{\theta}_t)^T\nabla\Phi(\theta_t)-\frac{1}{2}\sqrt{N} L\alpha^2_t ||g(\hat{\theta}_t)||^2 \nonumber
\end{align}
where $g(\hat{\theta}_t)=[\hat{\nabla}_{1} u_{1} (\theta_{1,t}, \hat{\theta}_{-1,t}),\cdots, \hat{\nabla}_{N} u_{N} (\theta_{N,t}, \hat{\theta}_{-N,t})]$ is the stochastic policy gradient vector defined as concatenation of individual gradients with respect to $\theta_{i,t}$ given the local copies $\hat{\theta}_{-i,t}$. By Lemma \ref{lem_unb} and potential game property, it holds, $\mathbb{E}[\hat{\nabla}_{i} u_{i} (\theta_{i,t}, \theta_{-i,t})| \ccalF_{t}]=\nabla_{i} u_{i} (\theta_{i,t}, \theta_{-i,t})$, for any agent $i \in \ccalN$. Together with Lipschitz continuity of (stochastic) policy gradients, we rewrite and separate the effect of local beliefs, by noting that  by $||\nabla\Phi(\theta_t)|| \le \sqrt{N}l$ and $ \mathbb{E}[||\theta_{j,t}- \hat{\theta}^i_{j,t}|| | \ccalF_t]= ||\theta_{j,t}- \hat{\theta}^i_{j,t}||$ as the parameters are inside the filtration $\ccalF_t$,
\begin{align}
     &\mathbb{E}[\Phi(\theta_{t+1})|\ccalF_t]-\Phi(\theta_{t})\nonumber\\
     &\ge \mathbb{E}[\alpha_t g(\hat{\theta}_t)^T\nabla\Phi(\theta_t)|\ccalF_t]-\frac{\sqrt{N}L}{2}\alpha^2_t \mathbb{E}[||g(\hat{\theta}_t)||^2|\ccalF_t] \nonumber
   \\
    &\ge  \alpha_t ||\nabla\Phi(\theta_t)||^2-\alpha_t \sqrt{N} l L \sum_{i \in \ccalN} \sum_{j \in \ccalN \setminus \{i\}} ||\theta_{j,t}- \hat{\theta}^i_{j,t}|| \nonumber\\
    & \hspace{0.5 cm }-\frac{2 L N^{3.5} R^2 V^2 }{(1-\gamma)^2(1-\sqrt{\gamma})^2}\alpha^2_t. \nonumber
\end{align}
  The right hand side above follows by Taylor expansion and the stochastic gradients updates. Note that it establishes \eqref{eq_pot_change}. Next, notice that $ \mathbb{E}[||g(\hat{\theta}_t)||^2|\ccalF_t] \le \frac{4 N^{3} R^2 V^2 }{(1-\gamma)^2(1-\sqrt{\gamma})^2}$ by Assumptions \ref{as_bound_rew}-\ref{as_bound_logp}, and therefore we have $\sum_{t=0}^\infty \frac{2 L N^{3.5} R^2 V^2 }{(1-\gamma)^2(1-\sqrt{\gamma})^2}\alpha^2_t = \sum_{t=0}^\infty \ccalO(t^{-2\beta}) < \infty$. 
Since the expected local errors have the rate $\mathbb{E}[||\theta_{j,t}- \hat{\theta}^i_{j,t}||]= \ccalO(t^{-\beta})$ by Lemma~\ref{lemma_local_conv} and Assumption~\ref{as_step}, we have the following by Fubini's theorem,
  \begin{equation}
      \mathbb{E} \bigg [ \sum_{t=0}^\infty \sum_{i \in \ccalN} \sum_{j \in \ccalN \setminus \{i\}} \alpha_t \sqrt{N} l L ||\theta_{j,t}- \hat{\theta}^i_{j,t}|| \bigg ]=\sum_{t=0}^\infty \ccalO(t^{-2\beta}).\nonumber
  \end{equation}
  Hence, we conclude that the infinite random sum is convergent in expectation, also implying convergence in probability. Thus, it converges almost surely as the sum is a non-decreasing monotonic sequence of random variables,
  \begin{equation}
      \sum_{t=0}^\infty \sum_{i \in \ccalN} \sum_{j \in \ccalN \setminus \{i\}} \alpha_t \sqrt{N} l L ||\theta_{j,t}- \hat{\theta}^i_{j,t}|| < \infty.\nonumber
  \end{equation}
   Thus, the sequence $\{ Z_t\}_{t\ge 0}$ adapted to  $\ccalF_t$ is a near super-martingale, integrable and non-negative $0 \le \mathbb{E} [Z_t] < \infty$  by Assumption \ref{as_bound_rew} (bounded rewards )and the potential property. We conclude that the random sum $\sum_{t=0}^\infty \alpha_t ||\nabla\Phi(\theta_t)||^2 < \infty $ is finite by Robbins-Siegmund Theorem \cite{robbins1971convergence}. 
\end{IEEEproof}


\blue{Compared to prior studies~\cite{eksin2017distributed,arefizadeh2019distributed,aydin2023policy,aydin2023policyL4DC,aydin2023networked}, 
we establish the potential change relying only on belief errors that vanish in 
expectation. The subsequent step differs from 
the single-agent case: the direction $g(\hat\theta_t)$ is evaluated at the 
agents' beliefs rather than at the true joint parameter, so it is not an 
unbiased estimate of $\nabla\Phi(\theta_t)$ and the first-order term does not 
collapse to $\alpha_t\|\nabla\Phi(\theta_t)\|^2$. Since this residual in \eqref{eq_pot_change} vanishes only in 
expectation, the optimality gap is a near-supermartingale rather than a 
supermartingale. We address this by showing that the expected sum of belief 
errors over time is convergent, and establish almost sure convergence through 
monotonicity together with the Robbins--Siegmund theorem.}
assumption of the prior studies.
\section{Simulations}
\label{sec:experiments}

\subsection{Dynamic Target Assignment}
\blue{
We consider a multi-agent dynamic target assignment problem in which $N$ agents move 
in the plane towards $K$ fixed targets and are penalized for selecting the same 
target. Let $y_k\in \reals^2$ denote the target 
positions for each $k \in \{1,\cdots,K\}$. The state of agent $i$ is its position $s_{i,t} \in \reals^2$, and 
the joint state is $s_t = (s_{1,t},\dots,s_{N,t})$. At each time $t$, each agent 
$i \in \ccalN$ selects a target index $a_{i,t} \in \ccalA := \{1,\dots,K\}$.

Each agent moves toward its selected target by at most a fixed step length, 
subject to additive noise,
\begin{equation} \label{eq_target_dyn}
s_{i,t+1} = s_{i,t} + \min\big\{v_i,\, \|y_{a_{i,t}} - s_{i,t}\|\big\}\,
\frac{y_{a_{i,t}} - s_{i,t}}{\|y_{a_{i,t}} - s_{i,t}\|} + \sigma_i \omega_{i,t},
\end{equation}
where $v_i > 0$ is the maximum distance agent $i$ travels in one period, 
$\sigma_i \ge 0$ is its noise level, and $\omega_{i,t} \sim \mathsf{N}(0, I_2)$ is 
independent across agents and time.
Since $s_{i,t+1}$ depends only on $(s_{i,t}, a_{i,t})$ and on agent $i$'s own 
noise, the transition kernel factorizes across agents. Agents are heterogeneous through 
the step lengths $v_i$, the noise levels $\sigma_i$, and their initial 
positions.

Agent $i$ is penalized for its distance to the selected target and for 
selecting a target that another agent also selects,
\begin{equation} \label{eq_target_reward}
r_i(s,a) =-\|s_{i} - y_{a_{i}}\| 
\;+ \sum_{j \in \ccalN \setminus \{i\}} c_{ij}\,\chi(a_{i}, a_{j}),
\end{equation}
where $\chi(a_i,a_j) := -\mathbb{1}_{\{a_i = a_j\}}$ is the pairwise 
collision penalty and $c_{ij} = c_{ji} \ge 0$ are symmetric interaction  weights. The distance term $\|s_{i,t} - y_{a_{i,t}}\| $ depends only on agent $i$'s own state and 
action, whereas the pairwise term couples the agents through their joint 
action. Therefore the function
$\phi(s,a) := -\sum_{j} \|s_{j} - y_{a_{j}}\| + \sum_{j<l} c_{jl}\chi(a_j,a_l)$ has the potential property by
summing the local terms and counting each interacting pair once at each state \cite{la2016potential}. 
Indeed, \(r_i(s,a)-\phi(s,a)\) is independent of \((s_i,a_i)\), establishing the potential property under unilateral deviations at every state \cite{la2016potential}. The corresponding potential value is
$\Phi(s) := \mathbb{E}[\sum_{t} \gamma^t \phi(s_t,a_t) \mid s_0 = s]$.

Agent $i$  maintains two parameters $\theta_i(k,1)$ and 
$\theta_i(k,2)$ for each target $k$ with distance feature 
$f_i(k) := \|s_i - y_k\|$. 
Its soft-max policy is
\begin{align} \label{eq_policy_target}
&\pi_i(a_i = k \mid s,\theta) \nonumber \\
&= \frac{\exp\big(\theta_i(k,1) 
+ \theta_i(k,2) f_i(k) - \theta^{\mathrm{mm}}_i(k)\big)}
{\sum_{k'=1}^{K}\exp\big(\theta_i(k',1) + \theta_i(k',2) f_i(k') 
- \theta^{\mathrm{mm}}_i(k')\big)},
\end{align}
where 
$\theta^{\mathrm{mm}}_i(k) := \widetilde{\max}_{j \in \ccalN \setminus \{i\}}
\{\theta_j(k,1) + \theta_j(k,2) f_j(k)\}$ and 
$\widetilde{\max}_{j \in \ccalN \setminus \{i\}}\{x_j\} := 
\log\big(\frac{1}{N-1}\sum_{j \in \ccalN \setminus \{i\}} \exp( x_j)\big)$ for 
$x_j \in \reals$ is the mellow-max operator, a differentiable approximation of 
the $\max$ operator proposed in \cite{asadi2017alternative}. 
It reduces the probability of selecting targets that are attractive to well-positioned competitors.

\noindent{\bf Numerical Setup:} 
$N=5$ agents, $K=5$ targets placed uniformly on the unit circle, i.e., 
$y_k = (\cos(2\pi k/K), \sin(2\pi k/K))$ for $k \in \{1,\dots,K\}$, and 
interaction weights $c_{ij}=1$ for all $i \neq j$. The step lengths $v_i$ and 
the noise levels $\sigma_i$ are drawn once per replication, uniformly from 
$[0.15,0.30]$ and $[0.005,0.015]$, respectively, so that agents differ in how 
far and how precisely they move in one period. The game starts with each agent 
at a position sampled from $\mathsf{N}(0,0.25^2 I_2)$, near the center of the 
target circle, where all targets are close to equidistant. Agents collect 
rewards with discount rate $\gamma=0.9$. We run each  algorithm 
with $T_f=20000$. Initial policy parameters are independently sampled from a 
normal distribution with $0$ mean and $0.1$ standard deviation, and initial 
beliefs satisfy $\hat\theta^i_{j,0} = \theta_{j,0} + \kappa\,\xi_{ij}$ with 
$\kappa = 0.5$ and $\xi_{ij} \sim \mathsf{N}(0,I)$ for $j \neq i$. Stepsizes have the form \(\alpha_0/\sqrt{t}\), with \(\alpha_0\in\{10,1,0.1\}\). We select \(\alpha_0\) using four separate replications and evaluate the selected configuration on disjoint seeds; all configurations select \(\alpha_0=0.1\). \footnote{The implementation and the scripts reproducing all figures are available at 
\url{https://github.com/sarperaydin/npg-mpg}.}

\begin{figure}[t] 
	\centering

    \includegraphics[width=.95\linewidth]{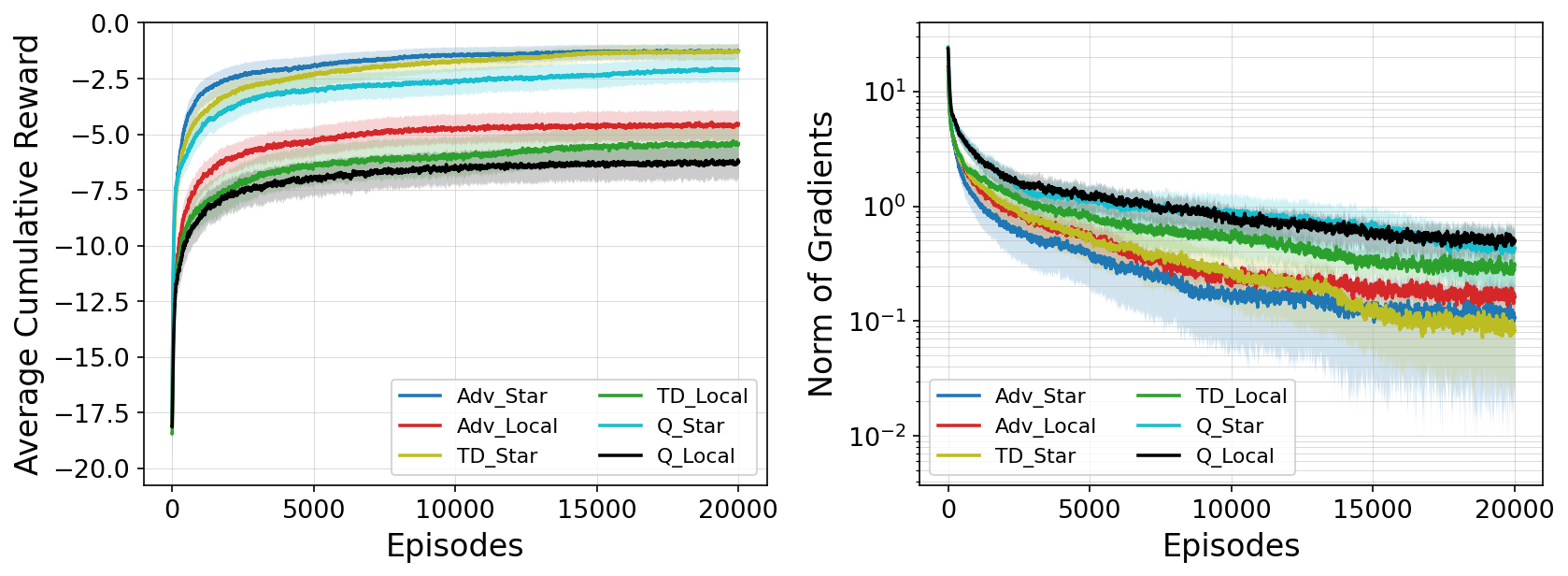}
         \vspace{-6pt}
\caption{\footnotesize Average cumulative rewards $\hat{R}_{i}$ 
at each episode (Left) and stochastic gradients 
$\frac{1}{N}\|\hat{\nabla}_i u_i(.)\|$ (Right) over $50$ replications. 
{\it Star} and {\it Local} indicate networked and local policies, 
respectively. We use the estimators Q, Adv, and TD for $\hat R_i$ defined in Algorithm \ref{alg_est_inner}. The lines show the 
exponential moving average with update rate of $0.05$, and shades show the 
$95\%$ confidence intervals. 
}
	\label{fig_netvsind}
\end{figure}
}

\subsubsection{Coupled versus Local Policies} \label{sec_cp_vs_loc}
\blue{We first isolate the effect of parameter coupling by comparing the networked policy in \eqref{eq_policy_target} with a local policy that depends only on agent $i$'s local state $s_i$ and its own parameters $\theta_i$. The local policy is obtained by removing the mellow-max coupling term from \eqref{eq_policy_target}. For the networked policy, we use an undirected time-varying communication network whose union forms a star over every four consecutive iterations.
For both policy classes, we evaluate the three policy-gradient estimators defined in Algorithm \ref{alg_est_inner}, 
based on the $Q$-function, advantage function, and temporal difference. These estimators are labeled $Q$, $Adv$, and $TD$ in Fig.~\ref{fig_netvsind}. We use \textit{Star} and \textit{Local} to denote the networked and local policy classes, respectively, resulting in six policy--estimator configurations.}

\begin{figure}[t] 
	\centering

	\begin{tabular}{ccc} 
	\includegraphics[width=.45\linewidth]{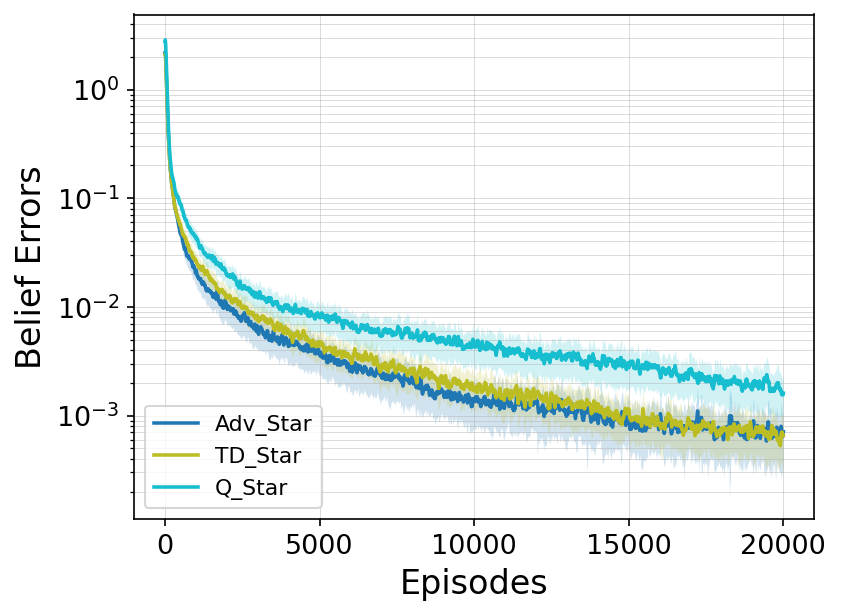} &
	\includegraphics[width=.45\linewidth]{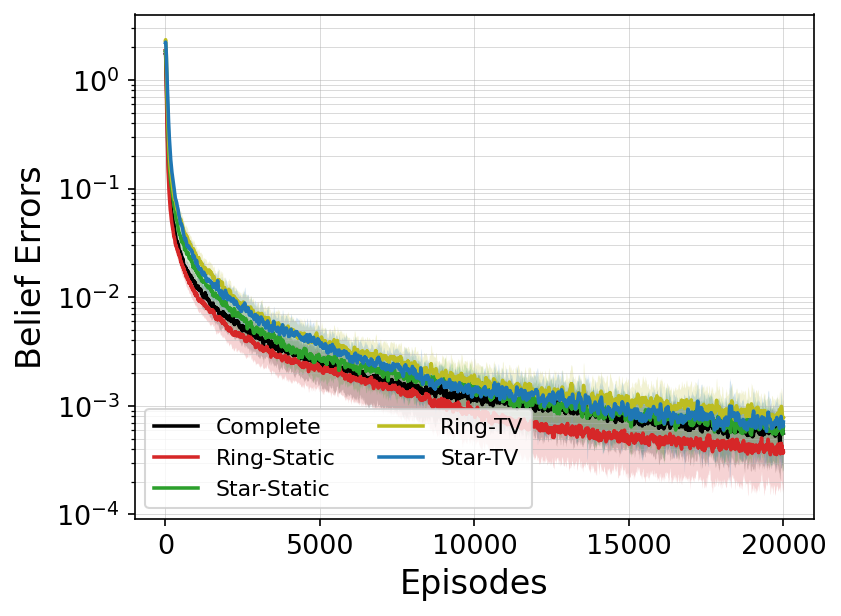}&
	\end{tabular}
     \vspace{-6pt}
	\caption{\footnotesize Local belief errors over $100$ replications on average $\frac{1}{N(N-1)}\sum_{i \in \ccalN} \sum_{j \in \ccalN \setminus \{i\}}||\theta_{i,t}-\hat{\theta}^j_{i,t} ||$ for (Left) Networked gradients with different reward estimations in a time-varying star-network (Right) Networked gradients with the advantage estimation in different communication network topologies. } 
	\label{fig_local_err}
\end{figure}

\blue{Coupled policies outperform local policies for all three estimators (Fig.~\ref{fig_netvsind}-Left). For both policy classes, the advantage and TD estimators perform similarly and outperform the \(Q\)-function estimator. Although zero reward would correspond to a collision-free assignment with every agent exactly at its target, process noise makes this value unattainable. The \(Q\)-function estimator also produces larger gradient norms and, consequently, larger parameter updates (Fig.~\ref{fig_netvsind}-Right).

To verify the assignments and coverage directly, for 
each replication we freeze the learned parameters, roll the joint policy for 
$H=30$ steps from $s_0$, and inspect the final step under two criteria. The 
first reads the joint action, requiring the $N$ selected targets to be distinct; 
the second reads the positions, counting a target as covered if some agent lies 
within $\epsilon = 0.05$ of it. Networked policies attain one-to-one assignments 
in $94\%$, $90\%$, and $74\%$ of replications under the Adv, TD, and $Q$ 
estimators, against $24\%$, $16\%$, and $6\%$ for local policies, and cover all 
five targets in $90\%$, $88\%$, and $66\%$ of replications against $20\%$, 
$12\%$, and $4\%$. Local policies therefore fail to coordinate on distinct 
targets rather than merely converging to inferior assignments.
Belief errors decrease with the gradient norms and approach \(10^{-3}\) by \(T_f=20{,}000\) (Fig.~\ref{fig_local_err}-Left). Adv and TD yield comparable errors, whereas \(Q\) produces larger errors because its larger gradients induce greater parameter changes between consensus updates.


\begin{figure}[t]
	\centering
	\includegraphics[width=.92\linewidth]{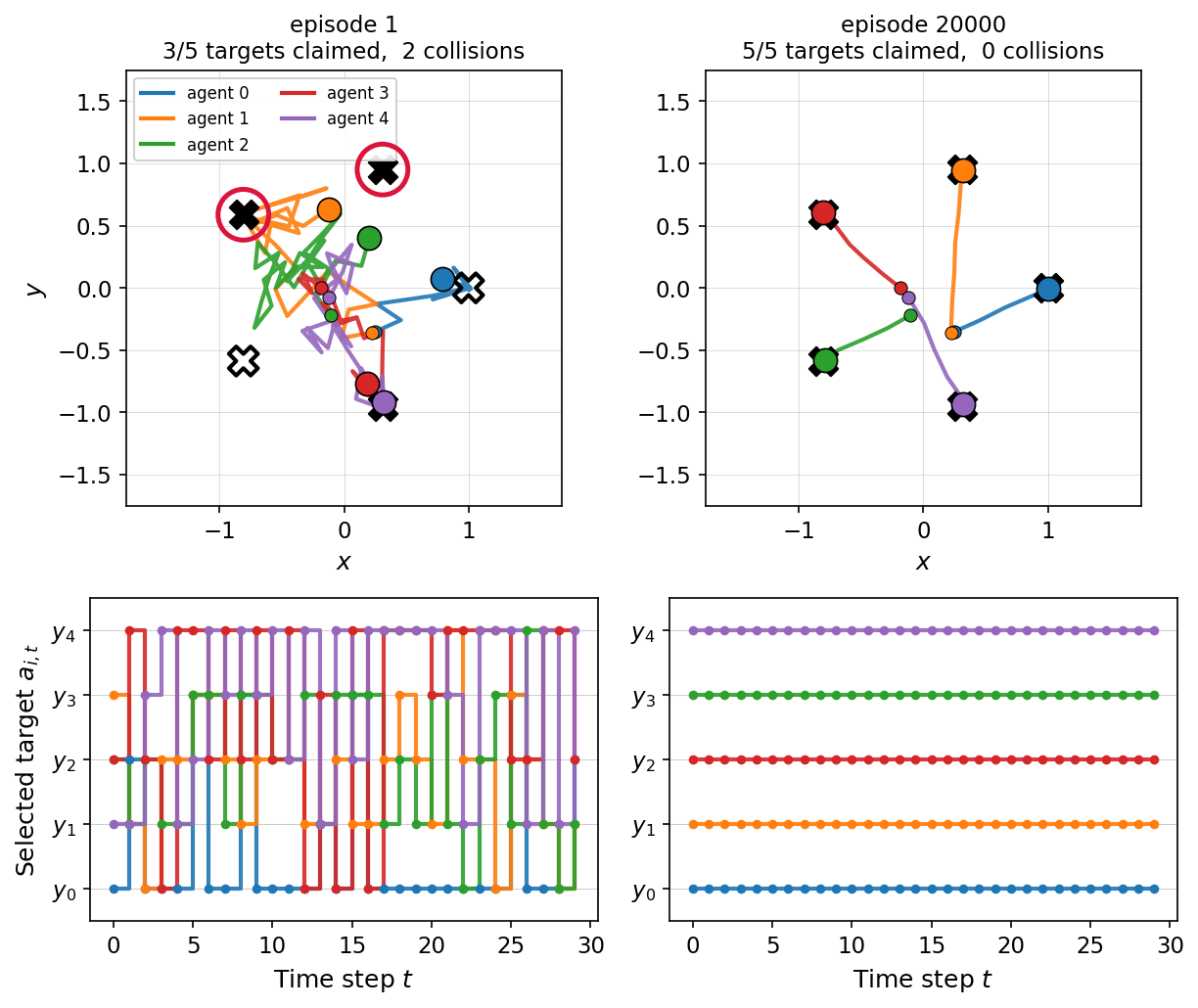}
         \vspace{-6pt}
	\caption{\footnotesize A single replication of the networked policy gradient 
play with the advantage estimator, before and after training. (Top) Agent 
trajectories over $30$ time steps under the policy at the corresponding 
episode, where small markers denote initial positions, large markers denote 
final positions, and crosses denote the targets. Hollow crosses indicate 
targets that no agent selects, and circled crosses indicate targets selected by 
more than one agent. (Bottom) The target $a_{i,t}$ selected by each agent along 
the same trajectory. At the beginning of training the agents switch targets 
$101$ times over the $30$ steps, leave two targets unclaimed and collide on a 
third; at the end of training each agent holds a distinct target for the entire 
trajectory. The average cumulative reward of this replication improves from 
$-13.57$ to $-0.68$ over the $T_f=20000$ episodes.}
	\label{fig_emergence}
\end{figure}

Fig.~\ref{fig_emergence} illustrates how parameter coupling resolves the assignment ambiguity. Because the agents begin near the center, distance alone does not distinguish among targets; coordination must emerge through the collision penalty. Training transforms frequent target switching and collisions into a stable one-to-one assignment in which every agent reaches a distinct target.
}


\subsubsection{Effect of Communication Topologies}
\blue{\begin{figure}[t] 
	\centering

    \includegraphics[width=.95\linewidth]{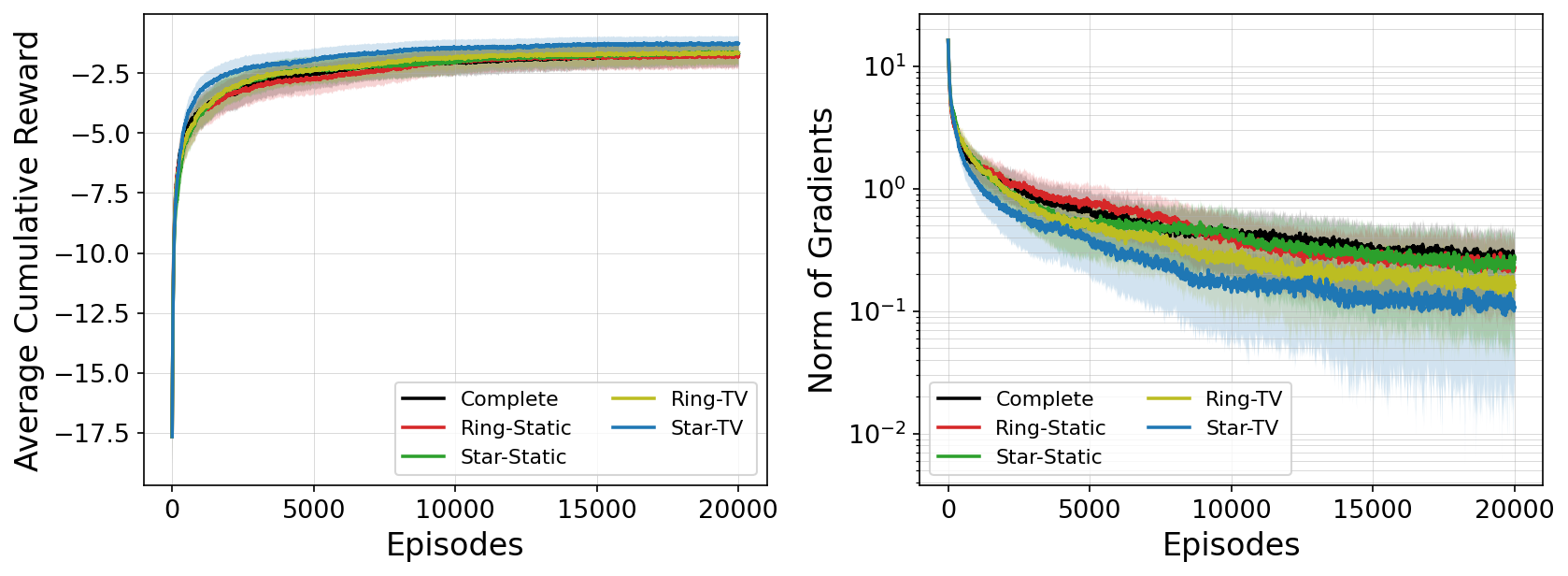}
         \vspace{-6pt}
	\caption{ \footnotesize Convergence results over 100 replications with different network topologies.  (Left) Average cumulative rewards with {\it Adv} estimator.  (Right)	Norms of stochastic Gradients $ \frac{1}{N} \sum_i || \hat{\nabla}_i u_i(.)||$. }\vspace{0pt}
	\label{fig_net_top}
\end{figure}
In Fig.~\ref{fig_net_top}, we compare the convergence of rewards and gradients 
for different communication network topologies including the
(complete) network, static and time-varying versions of the ring and star 
networks. 
All topologies satisfy Assumption \ref{as_connect}, but differ in 
the communication they consume per iteration. The complete network requires 
$N(N-1)$ transmissions, the static ring and star require $N$ and $2(N-1)$, 
while the time-varying ring and star require only one and two, respectively. 

The complete network learns slightly faster initially, but this gap disappears over time (Fig.~\ref{fig_net_top}-Left). In paired comparisons using common seeds, none of the four sparse topologies differs significantly from the complete network at the \(95\%\) level. Belief errors and gradient norms are similarly close across topologies (Figs.~\ref{fig_local_err} and \ref{fig_net_top}). Thus, additional communication density provides no measurable performance benefit in this problem: exchanging parameters matters, but frequent exchange does not.


The comparison indicates that the additional communication of the denser 
topologies is not converted into performance in this game. 
In contrast, the local policies, which communicate nothing, remain far behind all of 
them. What separates the algorithms is whether agents 
exchange their parameters at all, rather than how frequently they do so.
}

\subsubsection{Comparison with Baseline Policy Classes}

\blue{We next compare the networked policy with three baselines that separate the effects of joint-state conditioning and parameter exchange: (i) the local policy, which uses only local state and parameters; (ii) an independent policy, which conditions on the joint state without communication; and (iii) MAPPO, which uses centralized training with decentralized execution.

The {\it independent} policy retains the mellow-max coupling but supplies 
$M$ parameters per agent privately, where each agent maintains a parameter block for every agent, 
$\theta_i(j,k,\cdot)$ for $j \in \ccalN$, and forms
\begin{align} \label{eq_policy_private}
z_i(k) =& \theta_i(i,k,1) + \theta_i(i,k,2) f_i(k) \nonumber \\
&- \widetilde{\max}_{j \in \ccalN \setminus \{i\}} 
\big\{\theta_i(j,k,1) + \theta_i(j,k,2) f_j(k)\big\}.
\end{align}
Unlike \eqref{eq_policy_target}, all scores in \eqref{eq_policy_private} use parameters maintained privately by agent \(i\). Thus, the policy conditions on the joint state, but involves no parameter exchange and satisfies \(\partial\pi_n/\partial\theta_i=0\) for \(n\neq i\).

We also compare against MAPPO \cite{yu2022surprising}, which follows centralized training with decentralized execution. We give all agents the stage potential \(\phi(s,a)\) as a common reward and train a centralized critic on the joint state. This constitutes an information-rich baseline because evaluating the reward and critic requires global information. The actors use local soft-max policies with two parameters per target and the same initialization as \eqref{eq_policy_target}. The critic is a two-layer multilayer perceptron with 64 \(\tanh\) units per layer. We otherwise follow the training procedure in \cite{yu2022surprising} with complete state information.


The four algorithms perform different amounts of environment interaction per 
parameter update, so we compare them against the cumulative number of 
environment steps rather than against the iteration counter (Fig.~\ref{fig_ctde}). One episode of 
Algorithm~\ref{alg_est_inner} consumes $\ccalT_1$ transitions to reach the 
sampled pair and a further $\ccalT_2 - \ccalT_1 + 1$ for each of the return and 
value estimates, which is approximately $48$ steps in expectation. MAPPO instead 
collects $24$ parallel trajectories of $40$ steps and performs ten epochs of 
updates on each batch, so each of its iterations costs $960$ steps. This is why 
its curve begins two orders of magnitude further.
MAPPO  produces a steep rise after ten epochs are taken on every batch and the evaluation grid is coarse over the first iterations.

\begin{figure}[t]
	\centering
	\includegraphics[width=.78\linewidth]{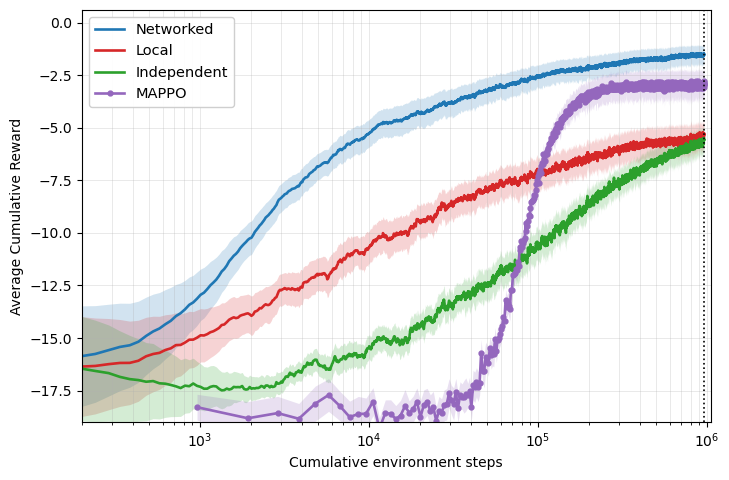}
         \vspace{-6pt}
\caption{\footnotesize Average cumulative rewards over $50$ replications 
against cumulative environment steps, for four algorithms (Networked, Local, Independent, MAPPO) run on the same set 
of problem instances. All four use the {\it Adv} estimator, and the first three share the sampling scheme in 
Algorithm~\ref{alg_est_inner}. All curves are 
truncated at a common budget of $0.96\times10^6$ steps. Lines 
show the exponential moving average with update rate $0.05$ and shades show the 
$95\%$ confidence intervals.}
	\label{fig_ctde}
\end{figure}

At the common budget the networked policy attains an average reward of $-1.53$, 
against $-2.77$ for MAPPO, $-5.65$ for the local policy and $-5.58$ for the 
independent policy. Its improvement over MAPPO is $1.43$, with a $95\%$ 
confidence interval of $[0.64, 2.22]$. Full target coverage shows the same 
ordering: the networked policy covers all five targets in $84\%$ of 
replications, against $54\%$ for MAPPO and $12\%$ and $10\%$ for the local and 
independent policies.


The independent policy offers no improvement over the lower-dimensional local policy despite observing the joint state and using \(N\) times as many parameters. Thus, joint-state conditioning alone does not explain the gain. Instead, the results point to parameter exchange as the key mechanism. The networked policy outperforms both decentralized baselines and the information-rich MAPPO baseline while requiring only two transmissions per iteration.


}

\subsection{Newsvendor Game}
\label{sec:news_vendorexperiments}
We consider a multi-agent newsvendor problem where
agents are vendors aiming to meet the total demand $D >0$ at each time $t$. Agents choose actions $a_{i} \in \{0,\cdots,N-1\}$ to determine the time when they supply the product to meet demand. 
In addition, they keep inventory $s_{i,t} \ge 0$ when their products are unsold. The joint state is the inventory of all the agents, i.e. $s_t=(s_{1,t}, \cdots s_{i,t},\cdots s_{N,t})$. In the game, agents need to coordinate their timings to match the demand and to reduce the cost of carrying inventory. Along these lines, we define the reward of each agent as below
\begin{equation}
    r_i= - c_o \frac{1}{N}\Big | \sum_{n \in \mathcal{N}} \mathbf{1}_{a_{n,t}=t \: mod \: N} D- D\Big |- c_s \frac{1}{N} \sum_{n \in \mathcal{N}} s_{n,t}
\end{equation}
where $c_o$ and $c_s$ are the unit opportunity cost and unit inventory cost, respectively. The indicator function $\mathbf{1}_{a_{n,t}=t \: mod \: N} \in \{0,1\}$ takes value as $1$ if the action of agent $n$ matches with the periodicity of the current time step $t$ which we model using the modulus of the total number of agents $N$. The structure of the reward implies agents are penalized in the case that the demand is not exactly matched, i.e., under-supplied or over-supplied. Further, the inventory incurs an additional cost. The inventory dynamics are as follows 
\begin{equation} \label{eq_inv_dy}
    s_{i,t}= \max \{ \mathbf{1}_{a_{i,t}=t \: mod \: N} D_, s_{i,t-1}\}- \frac{\mathbf{1}_{a_{i,t}=t \: mod \: N}}{\sum_{n \in \mathcal{N}} \mathbf{1}_{a_{n,t}=t \: mod \: N}} D.
\end{equation}
According to the inventory dynamics in \eqref{eq_inv_dy}, the inventory level for each agent $i$ changes only if agent $i$ attempts to sell a product at the current time, i.e., its action is equal to the periodicity of the time step $\mathbf{1}_{a_{i,t}=t \: mod \: N}=1$. 
If $\mathbf{1}_{a_{i,t}=t \: mod \: N}=1$, the inventory level of each agent $i$ increases to the demand level $D$ at the given time when the previous inventory level is not enough to satisfy demand. Assuming agents share the volume of demand equally with other agents, the inventory level $s_{i,t}$ of each agent $i$ only reduces with the amount of sold products inversely proportional to the total number of agents attempting to sell at the current time. We obtain the current inventory level in \eqref{eq_inv_dy} by subtracting the amount of products sold from the current inventory level. Otherwise, the inventory level stays the same.  

This game is an instance of Markov potential games with each agent having identical rewards, i.e. $r_i=r_j$ for all $i\in \ccalN$ and $j\in \ccalN$. 
%

\noindent{\bf Policy Parametrization:} Agent $i\in \ccalN$ maintains two parameters $\theta_{i}(k,1)$ and $\theta_{i}(k,2)$ per possible action, i.e., $k\in \{0,1,\dots,N-1\}$. We assume agents use a soft-max function as a probability distribution to sample from $N$ possible actions given the joint set of parameters and individual states,
\begin{align} \label{eq_policy}
    &\pi_i(a_{i}=k|s, \theta)= \frac{\exp\big( \theta_{i}(k,1)+ \theta_{i}(k,2) s_{i} - \tilde{\theta}_j(k)\big)} {\sum_{k'=0}^{N-1} \exp\big( \theta_{i}(k',1)+ \theta_{i}(k',2) s_{i} -\tilde{\theta}_j(k'))}
\end{align}
where $\tilde{\theta}_j(k):=\widetilde{\max}_{j \in \mathcal{N} \setminus \{i\} } \{{\theta}_{j}(k,1)+ {\theta}_{j}(k,2) s_{j} \}\big)$ and $\widetilde{\max}_{j \in \mathcal{N} \setminus \{i\} } \{  x_j \} := \log (\frac{1}{N-1} \sum_{j \in \mathcal{N} \setminus \{i\}} \exp{(x_j))}$ for $x_j \in \reals$ is the mellow-max operator which is a differentiable approximation of the $\max$ operator, proposed in \cite{asadi2017alternative}. 

\noindent{\bf Numerical Setup:} We set the game parameter values as follows: $N=5$, $D=1$, $c_o=0.3$, and $c_s=0.1$. The game starts with each agent having inventory, $s_{i,0}=1$ for $i\in\ccalN$. Agents collect rewards with discount rate $\gamma=0.95$. We run each instance of the algorithm with $T_f=2000$. Initial policy parameters are independently sampled from a normal distribution with $0$ mean and $0.3$ standard deviation.

\subsubsection{Coupled versus Independent Policies}

We compare the coupled policies in \eqref{eq_policy} against independent policies in which agents have policies that only depend on their own policy parameters. We remove the mellow-max terms in \eqref{eq_policy} to obtain the independent policies. For the networked policy we consider an undirected time-varying network whose union of edges form a star network in every five iterations.

\begin{figure}[t] 
	\centering
	\begin{tabular}{cc}
	\includegraphics[width=.45\linewidth]{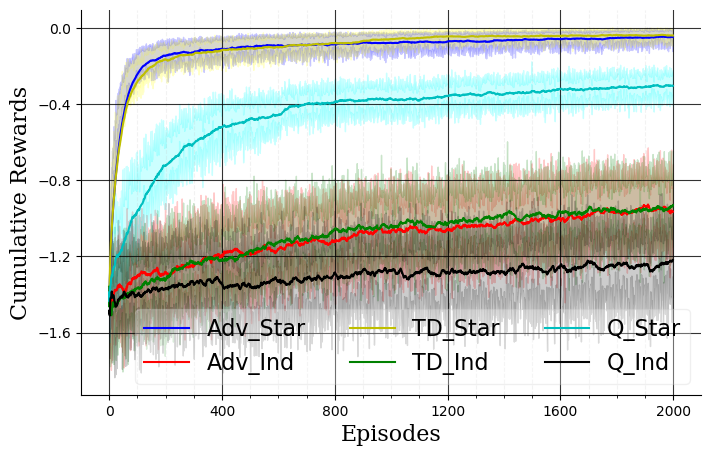} &
	\includegraphics[width=.45\linewidth]{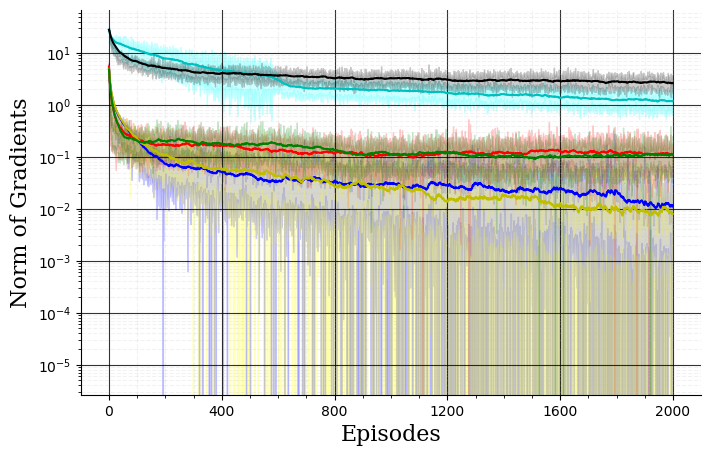}
	\end{tabular}
	\caption{ \footnotesize 
Average cumulative rewards $\hat{R}_{i} $ at each episode. (Left) and stochastic Gradients $ \frac{1}{N} || \hat{\nabla}_i u_i(.)||$ (Right) over 100 replications. {\it Star} and {\it Ind} indicates networked and independent policies, respectively. We use the estimators Q, Adv, and TD estimators. The lines show the exponential moving average with update rate of $0.05$, and shades show the 95\% confidence intervals. For each algorithm, the reported results correspond to the best-performing configuration in terms of final accumulated rewards, selected among stepsizes with initial magnitudes of orders $10^1$, $10^0$, and $10^{-1}$, following a diminishing rate of $1/\sqrt{t}$ for over time for each $t>0$.} 
	\label{fig_netvsind2}
\end{figure}

Fig. \ref{fig_netvsind2} shows the accumulated rewards and gradients over 100 replications for the networked policies versus the independent policies. 
In Fig. \ref{fig_netvsind2}-(Left), the coupled policies perform better than independent policies in collecting rewards within the $95 \%$ confidence interval. The most successful versions for both types of policies use either the advantage function or the temporal difference for $\hat R_i$. The two estimators' performances are close to each other throughout the iterations. We observe that the networked policies with the advantage function and the TD estimations converge near the optimal objective value $0$. 
We observe a similar performance pattern between the estimators $\hat R_i$ for independent policy gradient. The independent policies exhibit wider confidence intervals than the networked policies, demonstrating that coupled policies not only improve performance but also generate more reliable results.

Fig. \ref{fig_netvsind2}-(Right) shows the norm of the estimated gradients over time. The gradient estimations using $Q$-functions have higher norm values which can lead to abrupt changes in the parameter values. As also evident from Fig. \ref{fig_netvsind2}-(Left), coupled policies perform better than independent policies especially when using advantage and TD estimators. 
\begin{figure}[t] 
	\centering
	\begin{tabular}{cc}
	\includegraphics[width=.45\linewidth]{c_AUG24.png} &
	\includegraphics[width=.45\linewidth]{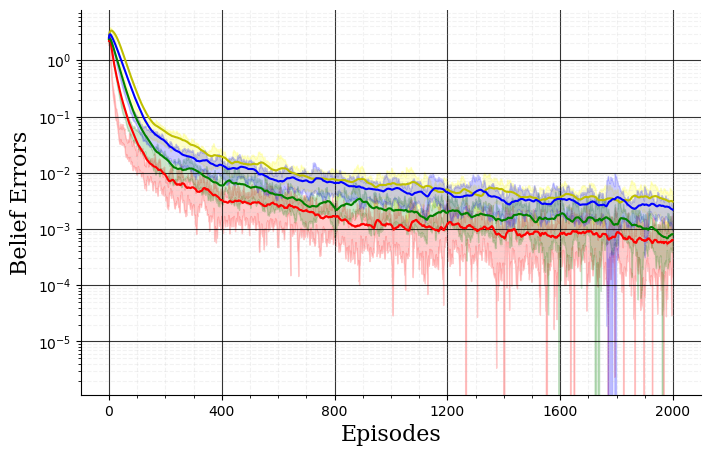}
	\end{tabular}
	\caption{\footnotesize Local belief errors over $100$ replications on average $\frac{1}{N(N-1)}\sum_{i \in \ccalN} \sum_{j \in \ccalN \setminus \{i\}}||\theta_{i,t}-\hat{\theta}^j_{i,t} ||$ for (Left) Networked gradients with different reward estimations in a time-varying star-network (Right) Networked gradients with the advantage estimation in different communication network topologies. } \label{fig_local_err2} \vspace{0pt}
\end{figure}

Fig. \ref{fig_local_err2}-(Left) illustrates the average belief errors among all the agents in networked policy gradients. The belief errors in all cases decrease over time and reach close to $0$ around the average norm value $10^{-2}$ by the final time $T_f=2000$. The belief errors show similar decreasing trend as the gradients. Indeed, we observe sharp changes in belief errors in steps that correspond to sharp gradient changes--see Fig. \ref{fig_netvsind2}-(Right).

\subsubsection{Effect of Communication Topologies}

\begin{figure}[t] 
	\centering
	\begin{tabular}{ccc} 
	\includegraphics[width=.45\linewidth]{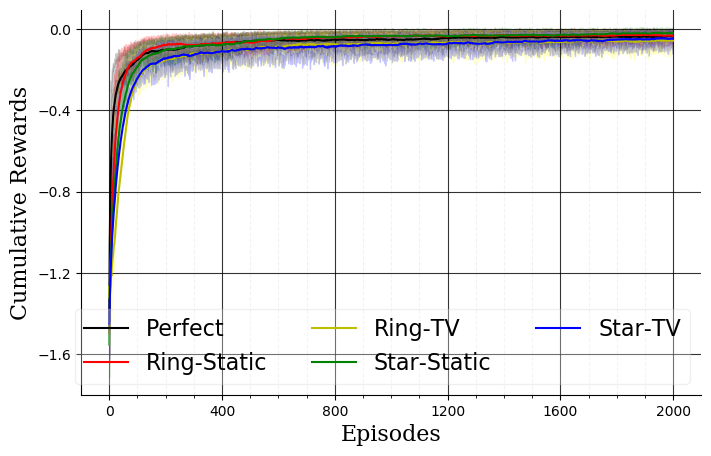} &
	\includegraphics[width=.45\linewidth]{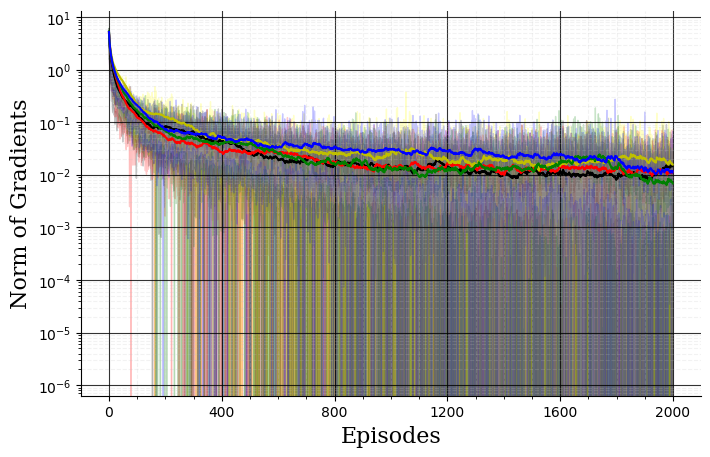}&
	\end{tabular}
	\caption{ \footnotesize Convergence results over 100 replications with different network topologies.  (Left) Average cumulative rewards $\hat{R}_{i} $ at each episode.  (Right)	Norms of stochastic Gradients $ \frac{1}{N} || \hat{\nabla}_i u_i(.)||$.}\vspace{0pt}
\label{fig_net_top2}
\end{figure}

In Fig.\ref{fig_net_top2}, we compare the convergence of rewards and gradients for different communication network topologies including the ``perfect'' (complete) network, static and time-varying versions of the ring and star networks. The difference is that the static networks satisfy strong connectivity among agents at each time step, while the time-varying networks only create strong connectivity in every $T=5$ time steps when we take the union of the edges over the $T=5$ steps. 

We do not observe a significant decline in performance switching from the perfect information case to other network types. As can be expected, the algorithm with perfect information has a slight lead in terms of accumulated rewards over the other network types until time $t\approx 400$ (Fig.\ref{fig_net_top2}-(Left)). However, the performance gap diminishes with the iterations such that all versions end close to the optimal value $0$. We observe similar patterns for belief errors (Fig.\ref{fig_local_err2}-(Right)) and gradients (Fig.\ref{fig_net_top2}-(Right)). 

\section{Conclusions}
\label{sec:conclusions}

In this study, we developed a networked policy gradient algorithm for Markov potential games with continuous and/or discrete state and action spaces. Agents use parametrized policies that depend on the policy parameters of other agents and maintain local estimates of these parameters
through consensus over a time-varying communication network. We established almost-sure convergence to stationary points of the potential value function and an expected iteration complexity of $\ccalO(\epsilon^{-2})$, matching standard stochastic-gradient methods. Numerical experiments on a dynamic multi-agent target-assignment problem and a multi-agent newsvendor game show that networked policies outperform communication-free local and independent policies, as well as a CTDE-based baseline. The results also show that this performance is retained over sparse time-varying networks.


\vspace{-5pt}
\appendix
\blue{We define the discounted state-visitation density
\begin{equation}\label{eq_dvis}
d^{\Pi_\theta}(s)=(1-\gamma)\sum_{t=0}^{\infty}\gamma^{t}\rho^\theta_t(s),
\end{equation}
where $\rho^\theta_t(s)$ is the density, with respect to the reference measure, 
of the state-action trajectory arriving at $s_t=s$ under the joint policy 
$\Pi_\theta$ from the initial state $s_0$,
\begin{equation}\label{eq_statedens}
\rho^\theta_t(s)=\prod_{h=0}^{t-1}\ccalP^{a_h}_{s_h,s_{h+1}}\prod_{h=0}^{t-1}\pi_\theta(a_h|s_h)
\Big|_{s_t=s}.
\end{equation}
The normalization $(1-\gamma)$ makes $d^{\Pi_\theta}$ a probability density, 
generalizing the discounted state distribution of \cite{sutton1999policy} and 
the state-occupancy measure of \cite{zhang2020global} to Borel state spaces. We 
further denote by
$\rho^\theta_{0:t+\tau}=\prod_{h=0}^{t+\tau-1}\ccalP^{a_h}_{s_h,s_{h+1}}
\prod_{h=0}^{t+\tau}\pi_\theta(a_h|s_h)$
the density of the full state-action trajectory $(s_0,a_0,\dots,s_{t+\tau},a_{t+\tau})$, 
which differs from \eqref{eq_statedens} in that the terminal action is included, 
as required when the reward at time $t+\tau$ is evaluated.}

\blue{\subsection{Proof of Lemma \ref{lem_pg_def}}\label{sec_app_pg_proof}
The Policy Gradient Theorem \cite{sutton1999policy} defines the policy gradient as
\begin{align}
    &\nabla_i u_i(\theta_i,\theta_{-i})
    \\
    &=\frac{1}{1-\gamma}\int_{\substack{a \in \mathcal{A}^N,\\ s \in \mathcal{S}}} Q_i^{\Pi_\theta} (s,a)\, d^{\Pi_\theta}(s)\, \nabla_i\pi_\theta(a|s) \, da \, ds,
 \end{align}   
where $d^{\Pi_\theta}$ is the discounted state-visitation density \eqref{eq_dvis}. Noting that $\pi_\theta(a|s)$ is the density of the joint policy $\Pi_\theta$, the log transformation on the gradient of the density, $\nabla_i\pi_\theta(a|s)$, yields
 \begin{align}   
     &\nabla_i u_i(\theta_i,\theta_{-i})\hspace{-2pt}\\
     &=\hspace{-2pt}\frac{1}{1-\gamma}\int_{\substack{a \in \mathcal{A}^N,\\ s \in \mathcal{S}}} Q_i^{\Pi_\theta}(s,a)\, d^{\Pi_\theta}(s)\, \pi_\theta(a|s)  \frac{\nabla_i\pi_\theta(a|s)}{\pi_\theta(a|s)}  \, da \, ds\\
    &=\frac{1}{1-\gamma}\int_{\substack{a \in \mathcal{A}^N,\\ s \in \mathcal{S}}} Q_i^{\Pi_\theta}(s,a)\,d^{\Pi_\theta}(s)\, \pi_\theta(a|s) \nabla_i \log \pi_\theta(a|s) \, da \, ds.
    \end{align}
Using the definition of networked policies in \eqref{eq_ind_pol}, and noting that $d^{\Pi_\theta}(s)\pi_\theta(a|s)$ is a probability density over state-action pairs,
    \begin{align}
    &=\frac{1}{1-\gamma}\int_{\substack{a \in \mathcal{A}^N,\\ s \in \mathcal{S}}} Q_i^{\Pi_\theta} (s,a)\, d^{\Pi_\theta}(s)\, \pi_\theta(a|s) \nonumber\\
    &\hspace{0.5 cm }\times \bigg( \sum_{n \in \ccalN} \nabla_i \log \pi_{n,\theta}(a_n|s) \bigg) \, da \, ds\\
    & = \frac{1}{1-\gamma} \: \mathbb{E}_{\Pi}\big[Q_i^{\Pi_\theta} (s,a)\sum_{n \in \ccalN} \nabla_i \log \pi_{n,\theta}(a_n|s) \big].
\end{align}
Moreover, any baseline function $b_i(s):\ccalS \rightarrow \reals $ which does not depend on the joint actions of agents $a \in \ccalA^N$ has the following property,
\begin{align} \label{eq_baseline_0}
    &\int_{a \in \mathcal{A}^N} b_i(s)  \nabla_i\pi_\theta(a|s) da = b_i(s) \nabla_i \bigg ( \int_{a \in \mathcal{A}^N} \pi_\theta(a|s)\, da \bigg )\\
    &= b_i(s) \nabla_i 1= 0.
\end{align}
Hence, the policy gradient of agent $i$ with a baseline function becomes
\begin{align}
    &\nabla_i u_i(\theta_i,\theta_{-i}) \nonumber \\
    &=\frac{1}{1-\gamma}\int_{\substack{a \in \mathcal{A}^N,\\ s \in \mathcal{S}}} ( Q_i^{\Pi_\theta}(s,a)- b_i(s))\, d^{\Pi_\theta}(s)\, \nabla_i\pi_\theta(a|s) \, da \, ds.
 \end{align}}

\subsection{Proof of Lemma \ref{lemma_lip}} \label{app_lem2}
We bound the norm of policy gradients in \eqref{eq_gradient} by Assumptions \ref{as_bound_rew}-\ref{as_bound_logp}, 
\begin{align}
 &|| \nabla_i u_i(\theta_i,\theta_{-i})|| \nonumber\\
 &   =\bigg|\bigg| \frac{1}{(1-\gamma)}\mathbb{E}\big[(Q_i^{\Pi_{\theta}} (s,a)-b_i(s)) \sum_{n \in \ccalN} \nabla_i \log  \pi_{n,\theta} (a_n |s)\big] \bigg| \bigg| \\
&\le \frac{1}{(1-\gamma)} \mathbb{E} \big[||(Q_i^{\Pi_{\theta}} (s,a)-b_i(s)) || .|| \sum_{n \in \ccalN} \nabla_i \log  \pi_{n,\theta} (a_n |s)||\big]. \label{eq_Fub_upb}
 \end{align}
 Since the norm of the difference $||(Q_i^{\Pi_{\theta}} (s,a)-b_i(s)) || \le \frac{2R}{1-\gamma}$ can be bounded using Assumptions \ref{as_bound_rew}-\ref{as_bound_logp}, we conclude boundedness of policy gradients
\begin{align}
|| \nabla_i u_i(\theta_i,\theta_{-i})||
\le \frac{2NRB}{(1-\gamma)^2}.
\end{align}


\blue{Using the definitions of the densities \eqref{eq_dvis}-\eqref{eq_statedens}, we expand the definition of the policy gradients, and drop the 
baseline term $b_i(s)$. Then, interchanging the 
integral with the summations by Fubini's theorem gives,}
\blue{\begin{align}
&\nabla_i u_i(\theta_i,\theta_{-i}) \nonumber \\ 
&=\frac{1}{1-\gamma}\int_{\substack{a_{0:t+\tau} \in (\mathcal{A}^N)^{t+\tau+1},\\s_{1:t+\tau} \in \mathcal{S}^{t+\tau}}} d^{\Pi_\theta}(s)\,  \pi_\theta(a|s) \nonumber \\ 
&\quad \times \bigg(\sum_{n \in \ccalN} \nabla_i \log \pi_{n,\theta}(a_n|s)\bigg) (Q_i^{\Pi_\theta} (s,a)-b_i(s)) \, da \, ds  \\
&=\sum_{n \in \ccalN} \sum_{t=0}^\infty \sum_{\tau=0}^\infty \gamma^{t+\tau}\int_{\substack{a_{0:t+\tau} \in (\mathcal{A}^N)^{t+\tau+1},\\s_{1:t+\tau} \in \mathcal{S}^{t+\tau}}}  r_{i}(s_{t+\tau},a_{t+\tau})\nonumber\\
&\quad\times\nabla_i \log \pi_{n,\theta}(a_{n,t}|s_t)\, \rho^\theta_{0:t+\tau} \, da_{0:t+\tau} \, ds_{1:t+\tau} \label{eq_pg_def_lip}
\end{align}}

 Further, we can remove the term coming from $b_i(s)$ in Eq. \eqref{eq_pg_def_lip}--see Lemma 1's proof. We now express the norm of the difference between the gradients defined at any two points $\theta^1,\theta^2 \in \reals^M$ as per \eqref{eq_pg_def_lip},
 \begin{align}
&|| \nabla_{i} u_{i} (\theta_{i}^1, \theta_{-i}^1) -\nabla_{i} u_{i} (\theta_{i}^{2}, \theta_{-i}^{2})||= \nonumber \\ \nonumber
& \Bigg| \Bigg |  \sum_{\substack{n \in \ccalN \\ t \in \naturals,  \tau \in \naturals}} \gamma^{t+\tau} \Bigg( \Bigg( \int_{\substack{a_{0:t+\tau} \in (\mathcal{A}^N)^{t+\tau+1},\\s_{1:t+\tau} \in \mathcal{S}^{t+\tau}}}  r_{i}(s_{t+\tau},a_{t+\tau})\\&\nonumber  \times (\nabla_i \log \pi_{n,\theta_1}({a_{n,t}}|s_t) - \nabla_i \log \pi_{n,\theta_2}({a_{n,t}}|s_t))  \nonumber\\
&\hspace{5 cm }\rho^{\theta_1}_{0:t+\tau}  da_{0:t+\tau} \, ds_{1:t+\tau}\Bigg ) \nonumber  \\ &+\Bigg (\int_{\substack{a_{0:t+\tau} \in (\mathcal{A}^N)^{t+\tau+1},\\s_{1:t+\tau} \in \mathcal{S}^{t+\tau}}}  r_{i}(s_{t+\tau},a_{t+\tau})\nabla_i \log \pi_{n,\theta_2}(a_{n,t}|s_t)\nonumber\\
&\hspace{3 cm}\times(\rho^{\theta_1}_{0:t+\tau}-\rho^{\theta_2}_{0:t+\tau})) da_{0:t+\tau} \, ds_{1:t+\tau} \Bigg ) \Bigg) \Bigg| \Bigg | \label{eq_Lip_norm_def}\\
&\le \sum_{\substack{n \in \ccalN \\ t \in \naturals,  \tau \in \naturals}}\gamma^{t+\tau} (||P_1||+||P_2||).
 \end{align}

For brevity, we denote 
$\sum_{\substack{n \in \mathcal{N},\, t \in \mathbb{N},\, \tau \in \mathbb{N}}}:
=\sum_{n \in \mathcal{N}} \sum_{t=0}^\infty \sum_{\tau=0}^\infty$ to represent the nested sum 
with a slight abuse of notation. We use $P_1$ and $P_2$ to refer to the two terms in parentheses \eqref{eq_Lip_norm_def}. Using triangle inequality, the norm of the first integral $P_1$ is bounded by Assumptions \ref{as_bound_rew}-\ref{as_bound_logp}
 \begin{align}
 & ||P_1||\le \nonumber\\
    & \int_{\substack{a_{0:t+\tau} \in (\mathcal{A}^N)^{t+\tau+1},\\s_{1:t+\tau} \in \mathcal{S}^{t+\tau}}}  |r_{i}(s_{t+\tau},a_{t+\tau})| .
     ||\nabla_i \log \pi_{n,\theta_1}({a_{n,t}}|s_t) \nonumber\\ &\hspace{0.5 cm}- \nabla_i \log \pi_{n,\theta_2}({a_{n,t}}|s_t)||  \: 
    \rho^{\theta_1}_{0:t+\tau}  da_{0:t+\tau} \, ds_{1:t+\tau} \\
    &\le R \ccalL || \theta^1-\theta^2|| \label{eq_bound_I1}.
 \end{align}
 
Next, we show Lipschitz continuity of the second integral, by defining the difference in probability {distributions} as the product of state transitions and policies,
\begin{align}
    &\rho^{\theta_1}_{0:t+\tau}-\rho^{\theta_2}_{0:t+\tau}\nonumber\\
    =& \prod_{h=0}^{t+\tau-1} \ccalP^{a_h}_{s_h,s_{h+1}} (\prod_{h=0}^{t+\tau} \pi_{n,\theta_{1}} (a_{n,h}|s_h) -\prod_{h=0}^{t+\tau} \pi_{n,\theta_{2}} (a_{n,h}|s_h)).
\end{align}

We bound the difference between two policies parametrized by $\theta^1,\theta^2 \in \reals^M$ with Taylor expansion,
\begin{align}\label{eq_pol_dif}
    &|\prod_{h=0}^{t+\tau} \pi_{n,\theta_{1}}(a_{n,h}|s_h) -\prod_{h=0}^{t+\tau} \pi_{n,\theta_{2}} (a_{n,h}|s_h)|\nonumber\\  \le& | (\theta^1-\theta^2)^T(\sum_{h'=0}^{t+\tau} \nabla_i \pi_{n,\Tilde{\theta}}(a_{n,h'}|s_{h'}) \prod^{t+\tau}_{\substack{h=0\\ h \neq h'}} \pi_{n,\Tilde{\theta}} (a_{n,h}|s_h))|,
\end{align}
where $\Tilde{\theta}=\varrho \theta^1+(1-\varrho)\theta^2$ is a convex combination of the points (vectors) $\theta^1,\theta^2 \in \reals^M$ for some $\varrho \in [0,1]$. The upper bound again follows from the triangle inequality, 
\begin{align}\label{eq_upper_for2}
    & \le || \theta^1-\theta^2|| \sum_{h'=0}^{t+\tau} || \nabla_i \log \pi_{n,\Tilde{\theta}} (a_{n,h'}|s_{h'})|| \prod^{t+\tau}_{h=0}\pi_{n,\Tilde{\theta}} (a_{n,h}|s_h)
\end{align}
where the term $\prod^{t+\tau}_{h=0}\pi_{n,\Tilde{\theta}} (a_{n,h}|s_h)$ is a product of proper probability densities. Inserting \eqref{eq_upper_for2} into the second integral in \eqref{eq_Lip_norm_def} gives 
\begin{align}
    & ||P_2||\le 
      || \theta^1-\theta^2|| (t+\tau+1)  R {V^2} \label{eq_bound_I2}.
\end{align}
where we use the following bound on the integral obtained using the Cauchy-Schwarz inequality
\begin{align}
    &\int_{\substack{a_{0:t+\tau} \in (\mathcal{A}^N)^{t+\tau+1},\\s_{1:t+\tau} \in \mathcal{S}^{t+\tau}}} ||\nabla_i \log \pi_{n,\theta_2}({a_{n,t}}|s_t) || \nonumber\\
    &\times || \nabla_i \log \pi_{n,\Tilde{\theta}} (a_{n,h'}|s_{h'})|| \nonumber\\
    &\times \prod^{t+\tau}_{h=0}\pi_{n,\Tilde{\theta}} (a_{n,h}|s_h) \prod_{h=0}^{t+\tau-1} \ccalP^{a_h}_{s_h,s_{h+1}}da_{0:t+\tau} \, ds_{1:t+\tau}  \le V^2.
\end{align} 
for any $\theta_2 \in \reals^M$ and $\Tilde{\theta}\in \reals^M$.

Combining the bounds \eqref{eq_bound_I1} and \eqref{eq_bound_I2} on the integrals concludes the Lipschitz continuity as follows,
\begin{align}
    &||\nabla_{i} u_{i} (\theta_{i}^1, \theta_{-i}^1) -\nabla_{i} u_{i} (\theta_{i}^{2}, \theta_{-i}^{2})||   \nonumber\\
    &\le \sum_{n \in \ccalN} \sum_{t=0}^\infty \sum_{\tau=0}^\infty \gamma^{t+\tau} R \bigg( \ccalL +{V^2} (t+\tau+1)\bigg) || \theta^1-\theta^2|| \\
    & \le N R\bigg(\frac{1}{(1-\gamma)^2} \ccalL +\frac{(1+\gamma) {V}^2}{(1-\gamma)^3 } \bigg) || \theta^1-\theta^2||
\end{align}
where the last bound is obtained using the relations $\sum_{t=0}^\infty \sum_{\tau=0}^\infty \gamma^{t+\tau}=(1/(1-\gamma))\sum_{t=0}^\infty  \gamma^t=1/(1-\gamma)^2$ and $\sum_{t=0}^\infty \sum_{\tau=0}^\infty \gamma^{t+\tau} (t+\tau+1)=\sum_{t=0}^\infty \gamma^t(-\gamma t +t+1)/ (1-\gamma)^2=(1+\gamma)/(1-\gamma)^3$.

\subsection{Proof of Lemma \ref{lem_unb}} \label{sec_app_lem_unb}

We state the expectation of the stochastic policy gradient, by taking expectations over the two different horizons $\ccalT_1$ and $\ccalT_2$ and the distribution of state-action pairs observed through these horizons consecutively. Similar to policy gradients, the stochastic gradients are bounded and integrable over the aforementioned random variables as the estimations of $Q$ and value functions, shown in Lemma \ref{lem_q_unb} with Assumptions \ref{as_bound_rew}-\ref{as_bound_logp}, so that the expectation operator can be separated and be written in any order, using Fubini's Theorem,
\begin{align}
&\mathbb{E}[\hat{\nabla}_i u_i(\theta_i,\theta_{-i})|\theta] \nonumber\\
=&\mathbb{E}_{\substack{\ccalT_1, s_{\ccalT_1},\\ a_{\ccalT_1}}}[\mathbb{E}_{\substack{\ccalT_2,s_{1:\ccalT_2},\\ a_{1:\ccalT_2} }}[\hat{\nabla}_i u_i(\theta_i,\theta_{-i})|s_{\ccalT_1}, a_{\ccalT_1},\theta] |\theta].
\end{align}
{Starting with the policy gradient without a baseline, }we indicate the value of the inner expectation using Lemma \ref{lem_q_unb} and the fact that the gradient of log-policy is independent of the given variables $\ccalT_2,s_{1:\ccalT_2}$ and $ a_{1:\ccalT_2}$,
\begin{align}
=&\mathbb{E}_{\substack{\ccalT_1, s_{\ccalT_1},\\ a_{\ccalT_1}}}\bigg[\mathbb{E}_{\substack{\ccalT_2,s_{1:\ccalT_2},\\a_{1:\ccalT_2} }}\big[\frac{1}{1-\gamma}\hat{Q}_i^{\Pi_{\theta}}( s_{\ccalT_1}, a_{\ccalT_1}) \nonumber\\ 
&\hspace{3 cm}\times \nabla_i \log \pi_\theta(a_{\ccalT_1}|s_{\ccalT_1})| s_{\ccalT_1}, a_{\ccalT_1}, \theta\big] \big |\theta\bigg]\nonumber\\
=&\mathbb{E}_{\substack{\ccalT_1, s_{\ccalT_1},\\ a_{\ccalT_1}}}\bigg[\frac{1}{1-\gamma}Q_i^{\Pi_{\theta}}(s,a)\nabla_i \log \pi_\theta(a|s)\big | \theta\bigg]\label{eq_outer_exp}
\end{align}

where we obtained the unbiased estimation of $Q$-values by independently sampled horizons. 
Since the norm of the gradient estimate is always bounded with the term for any state and action pair of $(s,a) \in \ccalS \times \ccalA^N$, its expectation is also bounded. We  further rewrite \eqref{eq_outer_exp} using indicator variables, and then by Fubini's Theorem, we take the summation over the infinite horizon out of the expectation,
\begin{align}
  &\mathbb{E}[\hat{\nabla}_i u_i(\theta_i,\theta_{-i})|\theta]\nonumber\\   =&\mathbb{E}_{\ccalT_1,s_{\ccalT_1}, a_{\ccalT_1}}\bigg[\sum_{t=0}^{\infty}\mathbb{1}_{ t = \ccalT_1} \frac{1}{1-\gamma}Q_i^{\Pi_{\theta}}(s,a) \nabla_i \log \pi_\theta(a|s)|\theta\bigg], \\
    =&\frac{1}{1-\gamma} \sum_{t=0}^{\infty}\mathbb{P}( t = \ccalT_1)\mathbb{E}_{s_{\ccalT_1}, a_{\ccalT_1}}[ Q_i^{\Pi_{\theta}}(s,a) \nabla_i \log \pi_\theta(a|s)|\theta],\\
    =&\sum_{t=0}^{\infty}\gamma^t \mathbb{E}_{s_{\ccalT_1}, a_{\ccalT_1}}[Q_i^{\Pi_{\theta}}(s,a) \nabla_i \log \pi_\theta(a|s)|\theta]
\end{align}

where we use the fact that $\ccalT_1$ follows a geometric distribution $P(t=\ccalT_1)=(1-\gamma) \gamma^{t}$. 
Next, we apply the definition of expectation and {again change the order of the summation and the integral using Fubini's Theorem},
\begin{align}
    &\mathbb{E}[\hat{\nabla}_i u_i(\theta_i,\theta_{-i})|\theta]\nonumber\\ =&\sum_{t=0}^{\infty}\gamma^t \int_{\substack{a \in \mathcal{A},\\ s \in \mathcal{S}}} \frac{1}{(1-\gamma)} Q_i^{\Pi_{\theta}}(s,a)\log \pi_\theta(a|s) \ccalP_{s,s',t}^{\pi_\theta} ds\: da,\\
    =&\int_{\substack{a \in \mathcal{A},\\ s \in \mathcal{S}}} \frac{1}{(1-\gamma)}\sum_{t=0}^{\infty}\gamma^t   Q_i^{\Pi_{\theta}}(s,a)\log \pi_\theta(a|s) \ccalP_{s,s',t}^{\pi_\theta} ds \: da.
\end{align}

Thus, we recover the definition of policy gradients by the fact  $d^{\Pi_\theta}(s)=\sum_{t=0}^\infty \gamma^{t} \ccalP_{s_0,s,t}^{\pi_\theta} $, and conclude that the stochastic gradient is unbiased,
\begin{align}
  &\mathbb{E}[\hat{\nabla}_i u_i(\theta_i,\theta_{-i})|\theta]  \nonumber\\=&\int_{\substack{a \in \mathcal{A},\\ s \in \mathcal{S}}} \frac{1}{(1-\gamma)} d^{\Pi_\theta}(s) Q_i^{\Pi_{\theta}}(s,a)\log \pi_\theta(a|s) ds da.
\end{align}

{Next, we also prove the remaining estimations of policy gradients, following a similar flow of proof. \begin{align}
=&\mathbb{E}_{\substack{\ccalT_1, s_{\ccalT_1},\\ a_{\ccalT_1}}}\bigg[\mathbb{E}_{\substack{\ccalT_2,s_{1:\ccalT_2},\\a_{1:\ccalT_2} }}\big[\frac{1}{1-\gamma}(\hat{Q}_i^{\Pi_{\theta}}( s_{\ccalT_1}, a_{\ccalT_1})-\hat{V}_i^{\Pi_{\theta}}( s_{\ccalT_1}))  \nonumber \\
&\hspace{2 cm}\times \nabla_i \log \pi_\theta(a_{\ccalT_1}|s_{\ccalT_1})| s_{\ccalT_1}, a_{\ccalT_1}, \theta\big]|\theta\bigg]\\
=&\mathbb{E}_{\substack{\ccalT_1, s_{\ccalT_1},\\ a_{\ccalT_1}}}\bigg[\mathbb{E}_{\substack{\ccalT_2,s_{1:\ccalT_2},\\a_{1:\ccalT_2} }}\big[\frac{1}{1-\gamma}(r_{i}(s_{\ccalT_1},a_{\ccalT_1})+\gamma \hat{V}_i^{\Pi_{\theta}}( s'_{\ccalT_1}) \nonumber \\
&\hspace{1.6 cm}-\hat{V}_i^{\Pi_{\theta}}( s_{\ccalT_1}))\nabla_i \log \pi_\theta(a_{\ccalT_1}|s_{\ccalT_1})| s_{\ccalT_1}, a_{\ccalT_1}, \theta\big]|\theta\bigg]\\
=&\mathbb{E}_{\substack{\ccalT_1, s_{\ccalT_1},\\ a_{\ccalT_1}}}\bigg[\frac{1}{1-\gamma}A_i^{\Pi_{\theta}}(s,a)\nabla_i \log \pi_\theta(a|s)|\theta\bigg]
\end{align}
Then, we again recover the definition of policy gradient definition as shown in Lemma \ref{lem_pg_def},
\begin{align}
=&\int_{\substack{a \in \mathcal{A},\\ s \in \mathcal{S}}} \frac{1}{(1-\gamma)}\sum_{t=0}^{\infty}\gamma^t   A_i^{\Pi_{\theta}}(s,a)\log \pi_\theta(a|s) \ccalP_{s,s',t}^{\pi_\theta} ds \: da \\
=&\int_{\substack{a \in \mathcal{A},\\ s \in \mathcal{S}}} \frac{1}{(1-\gamma)}\sum_{t=0}^{\infty}\gamma^t   Q_i^{\Pi_{\theta}}(s,a)\log \pi_\theta(a|s) \ccalP_{s,s',t}^{\pi_\theta} ds \: da.
\end{align}

}

\subsection{Unbiasedness of $\hat{Q}$ and $\hat{V}$}

 \begin{lemma} \label{lem_q_unb}
 The estimates $\hat{Q}_i^{\Pi_{\theta}}$ and ${\hat{V}_i^{\Pi_{\theta}}}$ for each agent $i \in \ccalN$ computed by Algorithm \ref{alg_DFP} is unbiased, $\mathbb{E}[\hat{Q}_i^{\Pi_{\theta}} (s,a)| s_0=s, a_{0}=a, \theta]=Q_i^{\Pi_{\theta}} (s,a)$ and {$\mathbb{E}[\hat{V}_i^{\Pi_{\theta}} (s)| s_0=s, \theta]=Q_i^{\Pi_{\theta}} (s)$. }
 \end{lemma}
 \begin{IEEEproof} 
 Using the definition of $\hat{Q}_i$ and {shifting the starting time from $\ccalT_1$ to $0$ without loss of generality}, we transform the finite sum of rewards over the sampled horizon length into an infinite sum,
 \begin{align}
 &\mathbb{E}[\hat{Q}_i^{\Pi_{\theta}} (s,a)| s_0=s, a_{0}=a, \theta] \nonumber\\
  =&\mathbb{E}[\sum_{t=0}^{\ccalT_2} \gamma^{t/2} r_{i}(s_t,a_t) | s_0=s, a_{0}=a, \theta]\\
 =&\mathbb{E}[\sum_{t=0}^{\infty} \mathbb{1}_{(0\le t \le \ccalT_2)} \gamma^{t/2} r_{i}(s_t,a_t) | s_0=s, a_{0}=a, \theta]\label{eq_est_def}.
 \end{align}
The absolute value of function sequence satisfies, $ |\sum_{t=0}^{\ccalT} \mathbb{1}_{(0\le t \le \ccalT_2)} \gamma^{t/2} r_{i}(s_t,a_t)| \le  \sum_{t=0}^{\ccalT} \gamma^{t/2} R$ by Assumption \ref{as_bound_rew}.  Then, since $\lim_{\ccalT \xrightarrow{} \infty} \sum_{t=0}^{\ccalT} \mathbb{E}( \gamma^{t/2} R)=\sum_{t=0}^{\infty} \gamma^{t/2} R < \infty$ exists and, is integrable, given $\gamma \in (0,1)$. Therefore, we rewrite \eqref{eq_est_def} with General Lebesgue Dominated Convergence Theorem,
\begin{align}
   &\mathbb{E}[\sum_{t=0}^{\infty} \mathbb{1}_{(0\le t \le \ccalT_2)} \gamma^{t/2} r_{i}(s_t,a_t) | s_0=s, a_{0}=a, \theta] \nonumber \\&=\sum_{t=0}^{\infty}\mathbb{E}[\mathbb{1}_{(0\le t \le \ccalT_2)} \gamma^{t/2} r_{i}(s_t,a_t) | s_0=s, a_{0}=a, \theta]
 \end{align}
The randomness stems from the stochastic joint policy and state transition together with the sampled horizon length. Next, we split the expectation of discounted state-action distribution and the distribution of random horizon lengths,
 \begin{align}
 =&\sum_{t=0}^{\infty}\mathbb{E}_{(s,a)}[\mathbb{E}_{\ccalT_2}(\mathbb{1}_{(0\le t \le \ccalT_2)}) \gamma^{t/2} r_{i}(s_t,a_t) | s_0=s, a_{0}=a, \theta]\\
 =&\sum_{t=0}^{\infty}\mathbb{E}_{(s,a)}[\gamma^{t} r_{i}(s_t,a_t) | s_0=s, a_{0}=a, \theta]
\end{align}

where we take expectation over the random variable $\ccalT_2 \sim Geom(1-\gamma^{0.5})$, such that {$ \mathbb{E}_{\ccalT_2}(\mathbb{1}_{(0\le t \le \ccalT_2)})=\mathbb{P} (\mathbb{1}_{(0\le t \le \ccalT_2)})= \gamma^{t/2}$}.
Again by General Lebesgue Dominated Convergence Theorem, we exchange the order of the expectation operator and summation to show the unbiasedness,

\begin{align}
    &= \mathbb{E}_{(s,a)}[\sum_{t=0}^{\infty}\gamma^{t} r_{i}(s_t,a_t) | s_0=s, a_{0}=a, \theta]= Q_i^{\Pi_{\theta}} (s,a).
\end{align}

{Following the same way of analysis, we can show unbiased estimate of value function of agent $i$ such that,
\begin{align}
&\mathbb{E}_{(s,a)}[\sum_{t=0}^{\infty} \mathbb{1}_{(0\le t \le \ccalT_2)} \gamma^{(t/2)} r_{i}(s_t,a_t) | s_0=s, \theta]\nonumber\\
=&\mathbb{E}_{(s,a)}[\hat{V}_i^{\Pi_{\theta}}(s) | s_0=s, \theta] =V_i^{\Pi_{\theta}}(s).
\end{align}
We conclude the lemma by also showing the unbiasedness of two forms of the advantage function estimation by using the unbiased estimates of $Q$-functions and value functions. First, we have the advantage estimation using separate estimations of $Q$ and value functions,
\begin{align}
&\mathbb{E}_{(s,a)}[\hat{A}^{\Pi_{\theta}}_{i} (s,a)| s_0=s, a_{0}=a, \theta ] \nonumber\\
=& \mathbb{E}_{(s,a)}[\hat{Q}_i^{\Pi_{\theta}}(s,a)-\hat{V}_i^{\Pi_{\theta}}(s)| s_0=s, a_{0}=a, \theta ]\\
=&Q_i^{\Pi_{\theta}}(s') -V_i^{\Pi_{\theta}}(s)=A^{\Pi_{\theta}}_i(s,a).
\end{align}
We also have the second estimation using value function estimations done at two sequential states $s \in \mathcal{S}$ and $s' \in \mathcal{S}$, where $s' \sim \mathbb{P}(.|s,\Tilde{a})$, and $a' \sim \Pi^{\theta}(.|s)$, 
\begin{align}
&\mathbb{E}_{(s,a)}[\widehat{TD}^{\Pi_{\theta}}_{i} (s,a) | s_0=s, a_0=a, \theta]\nonumber\\
=& \mathbb{E}_{(s,a)}[r_{i}(s,a)+
+\gamma\hat{V}_i^{\Pi_{\theta}}(s')-\hat{V}_i^{\Pi_{\theta}}(s)| s_0=s, \theta]\\
=&Q_i^{\Pi_{\theta}}(s,a) -V_i^{\Pi_{\theta}}(s)=A^{\Pi_{\theta}}_i(s,a).
\end{align}}
\end{IEEEproof}

\subsection{Proof of Theorem \ref{thm_as_con}} \label{ap_proof_as}

Since the stepsize is non-summable and the random sum $\sum_{t=1}^\infty \alpha_t ||\nabla\Phi(\theta_{t})||^2 < \infty$ is finite by Lemma \ref{lem_iter}, we obtain,
    \begin{equation} \label{eq_lim_inf}
        \liminf_{t \rightarrow \infty} ||\nabla\Phi(\theta_t)|| =0.
\end{equation}
In the remaining part, we will show that the limit supremum of the norm of gradients also goes to $0$, using non-summable stepsizes via a contradiction argument.
The result \eqref{eq_lim_inf} implies that there exists a sub-sequence, where the norm of the gradients have an upper bound $\epsilon/3>0$,
$    \ccalH_1=\{  \theta_t \in \reals^M: ||\nabla\Phi(\theta_t)|| \le \epsilon/3\}.$
Then, suppose that $\limsup_{t \rightarrow \infty} ||\nabla\Phi(\theta_t)|| > 2\epsilon/3$, which also implies the existence of a set below for some random realization of $\theta$ with {a positive probability},
%
     $   \ccalH_2=\{ \theta_t \in \reals^M: ||\nabla\Phi(\theta_t)|| \ge 2\epsilon/3\}.$

Since the gradient $||\nabla\Phi(\theta)|| $ is (Lipschitz) continuous, both sets are closed, leading to a well-defined Euclidian distance function between two sets,
$
    D(\ccalH_1,\ccalH_2)= \inf_{\theta^1 \in \ccalH_1} \inf_{\theta^2 \in \ccalH_2} || \theta^1-\theta^2||.
$

Note that these two sets are disjoint and their cardinality is infinite, which also ensures that the distance is always positive, $D(\ccalH_1,\ccalH_2) \ge \epsilon/3L>0$ by the reverse triangle inequality and Lipschitz continuity of the gradients (Lemma \ref{lemma_lip}). Thus, there exists a subsequence of iterations defined by the set
%
       $  \ccalH_3=\{ \theta_t \in \reals^M:  \epsilon/3 \le ||\nabla\Phi(\theta_t)|| \le 2\epsilon/3\}.$
%
Let $t_{1,\iota}$ and $t_{2,\iota}$ be the time steps that the iterates leave the set $\ccalH_1$ and enter the set $\ccalH_2$ respectively. We denote the time stamp of this event with $\iota \in \naturals^+$ for each occurrence. By Assumption \ref{as_step} and the fact ${D(\ccalH_1,\ccalH_2) \ge \epsilon/3L>0}$, there exists a nonempty set of iterates belonging to the set $\ccalH_{3,\iota}=\{ t \in \naturals : t_{1,\iota} < t < t_{2,\iota}\}$ after long enough finite time $t_{1,\iota}$ for some $\iota \in \naturals^+$. This implies $\ccalH_3= \cup_{\iota=1}^\infty \ccalH_{3,\iota}$ and $|\ccalH_3|= \infty$. Then, we bound the difference of iterates from below, 
\begin{align}
   &\sum_{\iota=1}^\infty \sum_{t=t_{1,\iota}}^{t_{2,\iota} } ||\theta_{t+1}-\theta_{t} || \ge \sum_{\iota=1}^\infty ||\theta_{t_{1,\iota}}-\theta_{t_{2,\iota}} ||\ge \sum_{\iota=1}^\infty D(\ccalH_1,\ccalH_2) \nonumber \\
    &\qquad  \ge \sum_{\iota=1}^\infty \frac{\epsilon}{3L}= \infty. \label{eq_sum_inf_dist}
\end{align}
The finiteness of the limit sum (Lemma \ref{lem_iter}) ensures
%
\begin{equation} \label{eq_sub_sums}
     \infty > \sum_{\iota=1}^\infty  \sum_{t=t_{1,\iota}}^{t_{2,\iota} } \alpha_t || \nabla\Phi(\theta_t) ||^2 \ge \sum_{\iota=1}^\infty \sum_{t=t_{1,\iota}}^{t_{2,\iota} } \alpha_t \frac{\epsilon^2}{9} .
\end{equation}
%
\blue{This implies $\sum_{\iota=1}^\infty \sum_{t=t_{1,\iota}}^{t_{2,\iota} } \alpha_t < \infty$. Since $\theta_{t+1}-\theta_t = \alpha_t g(\hat{\theta}_t)$, we bound the expected step length directly. Assumption \ref{as_bound_rew} gives the pathwise bound $|\hat{R}_i| \le 2R/(1-\sqrt{\gamma})$ for each of the estimators in Algorithm \ref{alg_est_inner}, so that $\mathbb{E}[||\hat{\nabla}_i u_i||] \le \frac{2R}{(1-\gamma)(1-\sqrt{\gamma})} \mathbb{E}[||\sum_{n \in \ccalN} \nabla_i \log \pi_{n,\theta}||] \le \frac{2NRB}{(1-\gamma)(1-\sqrt{\gamma})}$ by \eqref{eq_def_sg} and Assumption \ref{as_bound_logp}. Summing over the agents, $\mathbb{E}[||g(\hat{\theta}_t)||] \le \sum_{i \in \ccalN} \mathbb{E}[||\hat{\nabla}_i u_i||] \le \frac{2N^2RB}{(1-\gamma)(1-\sqrt{\gamma})}$. Together with \eqref{eq_sub_sums}, this yields
\begin{align} \label{eq_exp_finite_diff}
    \sum_{\iota=1}^\infty \sum_{t=t_{1,\iota}}^{t_{2,\iota} } \mathbb{E} [||\theta_{t+1}-\theta_{t} ||] 
    = \sum_{\iota=1}^\infty \sum_{t=t_{1,\iota}}^{t_{2,\iota} } \alpha_t \mathbb{E} [||g(\hat{\theta}_t) ||] \nonumber \\
    \le  \frac{2N^2RB}{(1-\gamma)(1-\sqrt{\gamma})}\sum_{\iota=1}^\infty \sum_{t=t_{1,\iota}}^{t_{2,\iota} } \alpha_t < \infty.
\end{align}}
%
%
The relation above shows that $\mathbb{E} [ \sum_{\iota=1}^\infty \sum_{t=t_{1,\iota}}^{t_{2,\iota} } ||\theta_{t+1}-\theta_{t} ||] < \infty$ by the monotone convergence theorem, as we exchange the order of expectation operator and the limit sum. This indicates $\sum_{\iota=1}^\infty \sum_{t=t_{1,\iota}}^{t_{2,\iota} } ||\theta_{t+1}-\theta_{t}|| < \infty $ due to Markov's inequality. However, this contradicts \eqref{eq_sum_inf_dist}. Hence, the claim that the limit supremum is bounded away from zero, i.e,  
 $\limsup_{t \rightarrow \infty} ||\nabla\Phi(\theta_t)|| > 2\epsilon/3$ is false for any $\epsilon >0$. Thus, we conclude that $\limsup_{t \rightarrow \infty} ||\nabla\Phi(\theta_t)||= 0$, which implies  $\lim_{t \rightarrow \infty} ||\nabla\Phi(\theta_t)||= 0$ almost surely, together with \eqref{eq_lim_inf}.

\subsection{Proof of Theorem \ref{thm_rate}} \label{app_proof_rate}
 Using \eqref{eq_pot_change}, we write the following inequality,
 \begin{align}
         \mathbb{E} [Z_{t+1}|\ccalF_t]&\le Z_{t}-\alpha_t ||\nabla\Phi(\theta_t)||^2  \\
         & \quad +\alpha_t L \sum_{i \in \ccalN} \sum_{j \in \ccalN \setminus \{i\}} ||\theta_{j,t}- \hat{\theta}^i_{j,t}||+\ccalO(t^{-2\beta}), \nonumber
 \end{align}
 where $Z_t=\Phi_{\max} -\Phi(\theta_t)$ is the optimality gap with respect to the parameter values $\theta_t$ at time $t$. Then, we rearrange the terms and sum over the time instances from $t-T$ to $t$ in order to bound the (unconditioned) expectations of the norm of the gradients, using $\mathbb{E} [||\theta_{j,\tau}- \hat{\theta}^i_{j,\tau}||]=\ccalO(\tau^{-\beta})$
 \begin{align}
     &\sum_{\tau=t-T}^t \mathbb{E}[||\nabla\Phi(\theta_\tau)||^2] \le \sum_{\tau=t-T}^t \frac{1}{\alpha_\tau} (\mathbb{E}[Z_{\tau}]-\mathbb{E}[Z_{\tau+1}])\nonumber\\
     & +L \sum_{\tau=t-T}^t \sum_{i \in \ccalN} \sum_{j \in \ccalN \setminus \{i\}} \mathbb{E} [||\theta_{j,\tau}- \hat{\theta}^i_{j,\tau}||]+\sum_{\tau=t-T}^t \ccalO(\tau^{-\beta}) \nonumber \\
      &= \sum_{\tau=t-T}^t \Bigg( \frac{1}{\alpha_\tau}-\frac{1}{\alpha_{\tau-1}}\Bigg )\mathbb{E}[Z_{\tau}] -\frac{1}{\alpha_{t}}\mathbb{E}[Z_{t+1}] \nonumber\\
      &\;\;  +\frac{1}{\alpha_{t-T-1}}\mathbb{E}[Z_{t-T}]+\sum_{\tau=t-T}^t \ccalO(\tau^{-\beta}),\label{eq_sum_grad_bound}
 \end{align}
where we obtain \eqref{eq_sum_grad_bound} by adding and subtracting $\mathbb{E}[Z_{t-T}]/\alpha_{t-T-1}$. Given Assumption \ref{as_bound_rew}, we have $0 \le Z_t \le \frac{2NR}{1-\gamma}$ and we expand the big-$\ccalO$ term with a constant $C>0$, leading to the following bound for \eqref{eq_sum_grad_bound},

\begin{align}
&\sum_{\tau=t-T}^t \mathbb{E}[||\nabla\Phi(\theta_\tau)||^2] \nonumber\\
&\le \hspace{-0.25 cm}\sum_{\tau=t-T}^t \Bigg( \frac{1}{\alpha_\tau}-\frac{1}{\alpha_{\tau-1}}\Bigg )\frac{2NR}{1-\gamma} +\frac{1}{\alpha_{t-T-1}}\frac{2NR}{1-\gamma}+C\sum_{\tau=t-T}^t \tau^{-\beta} \nonumber \\
&\le \frac{1}{\alpha_{t}}\frac{2NR}{1-\gamma}+C\sum_{\tau=t-T}^t \tau^{-\beta}.
\label{eq_sum_removed}
\end{align}
The last bound in \eqref{eq_sum_removed} follows from the removal of the nonpositive term $-\frac{1}{\alpha_{t}}\mathbb{E}[Z_{t+1}]$ resulting in the cancellation of the telescoping sum, and the replacing the value $\frac{1}{\alpha_t}=\Theta(t^\beta)$,
\begin{align}
    &\sum_{\tau=t-T}^t \mathbb{E}[||\nabla\Phi(\theta_\tau)||^2] \le O\bigg(t^\beta \frac{2NR}{1-\gamma}+ C(t^{1-\beta}-(t-T)^{1-\beta})\bigg), \nonumber
\end{align}
where we again bound the sum $C\sum_{\tau=t-T}^t \tau^{-\beta}=\ccalO(C(t^{1-\beta}-(t-T)^{1-\beta})$ with the corresponding definite integral $C\int_{t-T}^t \tau^{-\beta} d\tau$ for any $\beta \in (0,1)$. If $T=t-1$ and, when the sum on the right-hand side is divided by $t$,
\begin{align}
     \frac{1}{t}&\sum_{\tau=1}^t \mathbb{E}[||\nabla\Phi(\theta_\tau)||^2] \le \ccalO(t^{-\Tilde{\beta}}). \nonumber
\end{align}
where $\Tilde{\beta}:=\min\{1-\beta, \beta\}$. Hence, we have the property $\mathbb{E}[||\nabla\Phi(\theta_\tau)||^2]> \epsilon$ for any $t< T_{\epsilon}$ by its definition, yielding 
\begin{equation}
      \epsilon <  \frac{1}{T_{\epsilon}-1}\sum_{\tau=1}^{T_{\epsilon}-1} \mathbb{E}[||\nabla\Phi(\theta_\tau)||^2] \le  \ccalO(T_{\epsilon}^{-\Tilde{\beta}}).
\end{equation}
Thus, it concludes $T_{\epsilon} = \ccalO(\epsilon^{-1/\Tilde{\beta}})$, and the result is obtained when $\beta\to 1/2^+$.

\bibliographystyle{IEEEtran}
\bibliography{bibliography}

\vspace{-18pt}
\begin{IEEEbiography}[{\includegraphics[width=1.05in,height=1.25in,clip,keepaspectratio]{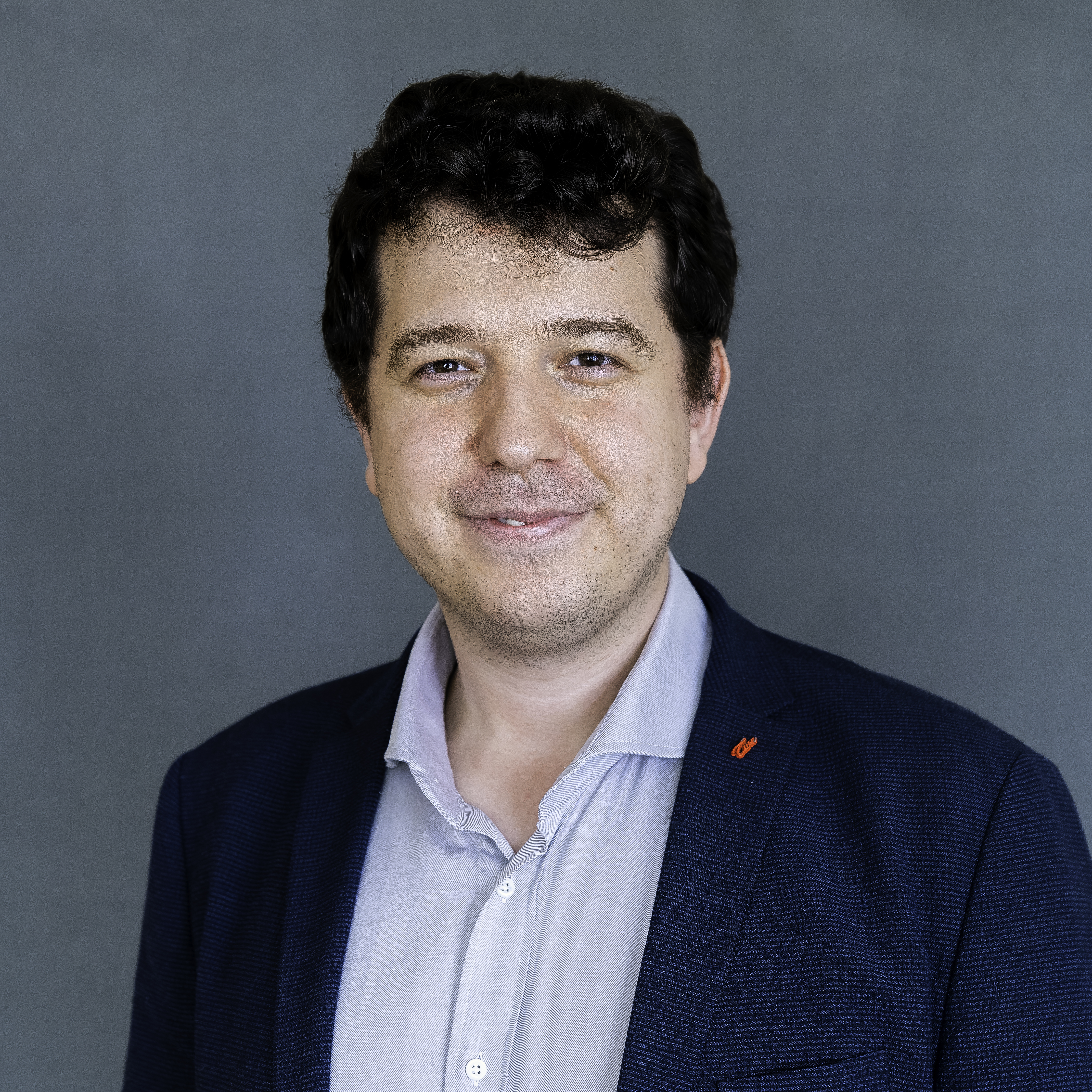}}] {Sarper Ayd\i n } (M'20) received his B.Sc. degree in Industrial Engineering from Bilkent University, Ankara, Turkey, in 2017. He completed his Ph.D. in Industrial Engineering in 2023 at Texas A\&M University. From 2023 to 2025, he was a postdoctoral scholar at Harvard University and Arizona State University. He is currently an Assistant Professor in the Department of Industrial and Management Systems Engineering at the University of South Florida. His research interests include decentralized and resilient algorithms for multi-agent systems.
\end{IEEEbiography}
\vspace{-18pt}
\begin{IEEEbiography}[{\includegraphics[width=1.05in,height=1.25in,clip,keepaspectratio]{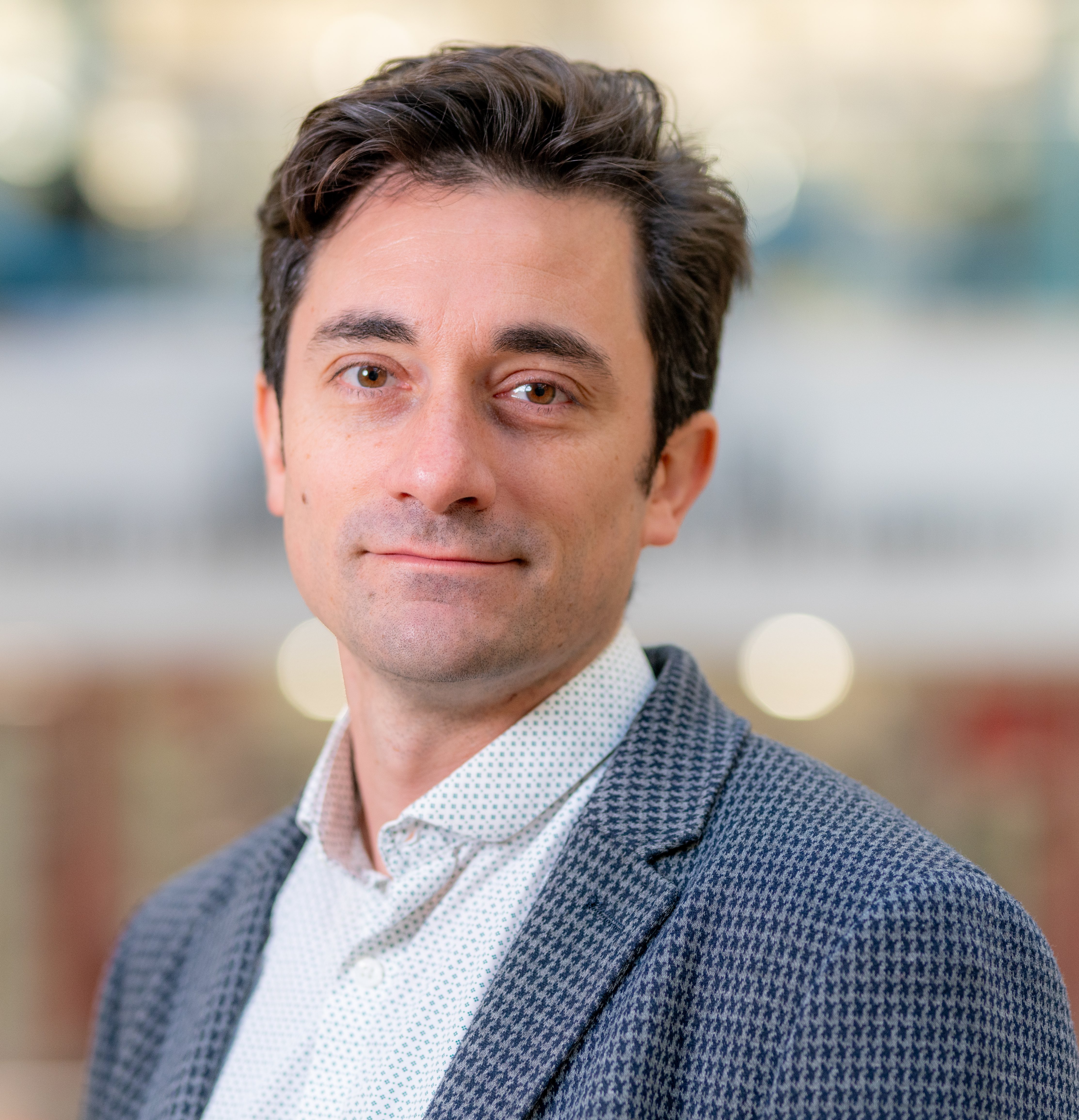}}] {Ceyhun Eksin} (M'11)
received the B.S. degree in control engineering from Istanbul Technical University, Istanbul, Turkey, in 2005, the M.S. degree in industrial engineering from Bo\u{g}azi\c{c}i University, Istanbul, Turkey in 2008, the M.A. degree in statistics from Wharton Statistics Department, University of Pennsylvania in 2015, and the Ph.D. degree in electrical and systems engineering from the Department of Electrical \& Systems Engineering, University of Pennsylvania in 2015. He was a postdoctoral researcher jointly affiliated with Schools of Biological Sciences, and Electrical and Computer Engineering, Georgia Institute of Technology. He is currently an associate professor in the Department of Industrial and Systems Engineering at Texas A\&M University, College Station, TX, USA. He is a recipient of NSF CAREER award in 2023. His research  focuses on distributed learning, game theory, and optimization with applications to autonomous teams, epidemics and energy systems.

\end{IEEEbiography}

\end{document}